\documentclass{JHEP3}
\usepackage{amssymb,epsf,graphicx}
\usepackage{amsmath}
\allowdisplaybreaks[1]
\newcommand{\sgn}{\mathop{\mathrm{sgn}}}

\def\Im{\mathop{\mathrm{Im}}}
\def\Re{\mathop{\mathrm{Re}}}

\def\state#1{\left|#1\right\rangle }
\def\L{\mathcal{L}}
\def\del{\partial}
\def\xp{x^{+}}
\def\yp{y^{+}}
\def\Vp{V^{+}}
\newcommand{\ordB}[0]{\genfrac{}{}{0pt}{2}{\star}{\star} }
\def\corr#1{\left\langle #1\right\rangle }

\title{Delays in Open String Field Theory}

\author{Frederik Beaujean\\ Max-Planck-Institut f\"ur Physik\\ 
F\"ohringer Ring 6, 80805 M\"unchen, Germany\\ 
E-mail: \email{beaujean@mpp.mpg.de}}

\author{Nicolas Moeller\\ Arnold-Sommerfeld-Center for Theoretical Physics\\ 
Department f\"ur Physik, Ludwig-Maximilians-Universit\"at M\"unchen\\
Theresienstra\ss e 37, 80333 M\"unchen, Germany\\ 
E-mail: \email{nicolas.moeller@physik.uni-muenchen.de}}

\abstract{We study the dynamics of light-like tachyon condensation in
  a linear dilaton background using level-truncated open string field
  theory. The equations of motion are found to be delay differential
  equations. This observation allows us to employ well-established
  mathematical methods that we briefly review.  At level zero, the
  equation of motion is of the so-called retarded type and a solution
  can be found very efficiently, even in the far light-cone future. At
  levels higher than zero however, the equations are not of the
  retarded type. We show that this implies the existence of
  exponentially growing modes in the non-perturbative vacuum, possibly
  rendering light-like rolling unstable. However, a brute force
  calculation using exponential series suggests that for the
  particular initial condition of the tachyon sitting in the false
  vacuum in the infinite light-cone past, the rolling is unaffected by
  the unstable modes and still converges to the non-perturbative
  vacuum, in agreement with the solution of Hellerman and Schnabl.
  Finally, we show that the growing modes introduce non-locality
  mixing present with future, and we are led to conjecture that in the
  infinite level limit, the non-locality in a light-like linear
  dilaton background is a discrete version of the smearing
  non-locality found in covariant open string field theory in flat
  space.}


\preprint{MPP-2009-194 \\ LMU-ASC 50/09}

\begin{document}

%%%%%%%%%%%%%%%%%%%%%%%%%%%%%%%%%%%%%%%%%%%%%%%%%%%%%%%%%%%%%%%%%%%%%%%%%%%%%%
\section{Introduction}
\label{s_intro}
%%%%%%%%%%%%%%%%%%%%%%%%%%%%%%%%%%%%%%%%%%%%%%%%%%%%%%%%%%%%%%%%%%%%%%%%%%%%%%

While the free action of open string field theory \cite{NNW} is local
in the sense that it involves not more than two derivatives of the
string field, it is well known that the interaction term of Witten's
string field theory \cite{Witten:1985cc} contains infinitely many
derivatives. Theories with more than two, but finitely many,
derivatives \cite{Ostrogradski,Eliezer:1989cr} suffer either from an
unbounded Hamiltonian or from ghosts\footnote{See, however,
  \cite{Bender:2007wu} for an example of quantization of a theory with
  four derivatives, with no ghost and bounded (but non-hermitian)
  Hamiltonian.}, but these instabilities do not necessarily survive in
the limit of infinitely many derivatives. Perhaps the simplest way to
see this is that the propagator in a theory with finitely many
derivatives is the inverse of a polynomial and therefore has poles,
some of them ghosts. In the limit of infinite number of derivatives,
however, the propagator becomes the inverse of a function that might
have only one zero, corresponding to a regular excitation. It might
even have no zero at all, like in $p$-adic string theory where the
propagator is an exponential (furthermore, this exponential propagator
renders all loop diagrams finite \cite{Minahan:2001pd}).

In cubic open string field theory, the form of the nonlocality is
universal, the higher derivatives of any field $\phi(x)$ always
appearing in the interaction as
\begin{equation}
\tilde{\phi}(x) \equiv K^{\Box} \phi(x),
\label{tilde} 
\end{equation}
where $K = \frac{3 \sqrt{3}}{4}$, and we use the signature $\eta_{\mu
  \nu} = {\rm diag}(-1,1,\ldots,1)$. In this case one can see
 the nonlocality explicitly because $\tilde{\phi}(x)$ is a smearing of
$\phi(x)$ as can be seen from the convolution formula
\cite{Brekke:1988dg}
\begin{equation}
e^{\beta \partial_x^2} \phi(x) = 
\frac{1}{2 \sqrt{\pi \beta}} \int_{-\infty}^\infty 
e^{-\frac{1}{4 \beta} (x-y)^2} \phi(y) dy, \qquad \beta >0.
\label{convolution}
\end{equation}
For a homogeneous time-dependent problem, one would take
$\tilde{\phi}(t)$ as the fundamental field and write $\phi(t) =
e^{\log(K) \partial_t^2} \tilde{\phi}(t)$, which can then be written
as a convolution as in Eq.~(\ref{convolution}). A consequence is that
the equation of motion of a homogeneous time-dependent string field
involves the string field not only at time $t$ but at {\em all} times,
both in the past and in the future of $t$. One can treat this kind of
equation either as a convolution equation using
Eq.~(\ref{convolution}), or as a differential equation of infinite
order. It must be understood, however, that such a differential
equation cannot be seen as a limit of finite-order differential
equation; in particular the initial value problem becomes different
when we have infinitely many derivatives (see \cite{Barnaby:2007ve}
for a rigorous discussion, and \cite{Calcagni:2007ef} which contains
some similar results).

A time-dependent equation of motion with infinitely many derivatives
which is of particular physical interest, is the equation describing
the decay of an unstable D-brane. A well-known problem is that, on the
one hand, a boundary conformal field theory (BCFT) analysis shows that
one should expect a monotonic decay of the tachyon down its potential
\cite{Sen:2002nu} and that the energy of the D-brane is converted into
very massive closed strings at rest, behaving like dust
\cite{Sen:2002in} (tachyon matter). On the other hand, numerical
solutions of string field theory \cite{Moeller:2002vx,Fujita:2003ex}
show a completely different behavior. Namely, the tachyon does reach
the non-perturbative vacuum, but it then continues further and starts
oscillating around it with diverging amplitude. Although the tachyon
can climb arbitrarily high up the potential, the energy conservation
is not violated because the kinetic energy can be negative. In fact,
it may seem that energy conservation obviously discards a monotonically
rolling tachyon that would stop at the local minimum of its potential.
However, this is not totally trivial because one could imagine that
the energy of the D-brane is somehow stored in the very high-order
derivatives of the tachyon, still allowing for a monotonic rolling.
But this is actually ruled out \cite{Yang:2002nm,Joukovskaya:2008cr}
because one can write the expression of the energy in an integral form
which makes it clear that, in fact, a monotonically rolling tachyon
cannot conserve energy. It is now believed that these ever-growing
oscillations are not catastrophic after all. For one thing the string
field is not a gauge-invariant observable; but more concretely, it was
shown in \cite{Coletti:2005zj}, that a field redefinition mapping the
cubic SFT action to the boundary SFT action, would also map the
oscillating solution to a well-behaved solution. More recently, it was
shown \cite{Kiermaier:2008qu} that the closed string boundary state
obtained from the rolling tachyon solution, coincides with the BCFT
boundary state.

It is interesting to investigate how this wild rolling changes if we
somehow couple the closed strings sector to the open string SFT
action. The most consistent way to do this would be to consider
open-closed string field theory \cite{Zwiebach:1997fe}; but solving
the equations of motion of the purely closed sector
\cite{Zwiebach:1992ie} involves a much higher level of difficulty
\cite{Belopolsky:1994bj,Moeller:2004yy,Yang:2005ep,Yang:2005iua,Yang:2005rx,
  Moeller:2006cv,Moeller:2006cw,Moeller:2007mu,Moeller:2008tb}. A
somewhat more manageable approach would be to consider a fixed closed
string background, but the SFT action becomes in general
non-polynomial in a generic closed background \cite{Zwiebach:1990qj},
hence also hard to solve. What can be done, however, is to minimally
couple gravity to the open SFT action. It has been shown, for
instance, that minimally coupling an open superstring tachyon to a FRW
metric tames the wild oscillations of the tachyon; and convergent
rolling tachyon solutions were found numerically
\cite{Joukovskaya:2008cr}. In \cite{Hel-Sch}, which was the motivation
for the present work, Hellerman and Schnabl considered open SFT in a
linear dilaton background. They chose a {\em light-like} dilaton
gradient and a string field depending only on the light-cone time
$\xp$. This is physically motivated because a bubble of true vacuum is
expected to expand at the speed of light \cite{Coleman:1977py}. If the
radius of the bubble is large enough, we can focus on one small patch
and approximate it by a plane, and we choose the light-light
coordinate $x^+$ (which we call light-cone time) to be orthogonal to
this plane. Moreover, with this ansatz important simplifications
occur. In particular, using the fact that $e^{X^+}$ is an exactly
marginal operator, Hellerman and Schnabl were able to use the results
of \cite{Schnabl:2007az,Kiermaier:2007ba} in order to prove that the
rolling tachyon asymptotes to the tachyon vacuum \cite{Schnabl:2005gv}
at large light-cone time. On a more explicit footing, Hellerman and
Schnabl also considered the SFT action truncated at level zero in
Siegel gauge (i.e. keeping only the tachyon). Here the light-cone
simplification manifests itself by changing the nature of the
non-locality. The non-locality of Eq.~(\ref{tilde}), which by virtue
of Eq.~(\ref{convolution}), involves the tachyon field at all times,
becomes simply the tachyon at some {\em retarded light-cone time}
$\phi(x^+-\gamma)$. The equation of motion for the tachyon is then
\begin{equation}
  \phi'(\xp) - \phi(\xp) = - K^3 \, \phi(\xp-\gamma)^2.
\label{TachEOMlev0}
\end{equation}
A numerical solution to this equation was worked out by Hellerman and
Schnabl. Their method, however, didn't allow them to go very far in
light-cone time, but enough to see convincingly that the tachyon
reaches the vacuum after oscillating around it with a decreasing
amplitude. In \cite{Barnaby}, Barnaby et al. considered the initial
value problem and the stability of light-like rolling in $p$-adic
string theory and in SFT at level zero. In particular, they were able
to numerically solve Eq.~(\ref{TachEOMlev0}) for a much larger
light-cone time interval. Their numerical method is based on the
diffusion equation
\cite{Calcagni:2007wy,Calcagni:2009tx,Calcagni:2009jb}. Although this
method can in principle be generalized to higher levels, it is hard to
do so in practice.

The motivation for our present paper, was to investigate further the
light-cone rolling in Siegel gauge by considering higher-level fields.
We will consider levels $(2,4)$, $(2,6)$, and $(4,8)$, where the notation
$(L,M)$ means that we are keeping fields up to level $L$ and
interactions up to total level $M$. Our results are three-fold.

Firstly, we realized that equations of the type (\ref{TachEOMlev0})
are known in the mathematics literature as {\em delay differential
  equations} (abbreviated DDEs). The most widely used method for
numerically solving DDEs is the {\em method of steps}. Using this
method, we show that solving Eq.~(\ref{TachEOMlev0}) numerically
becomes surprisingly easy. Moreover, the generalisation to higher
levels is straightforward.

Secondly, we show that when we include higher-level fields, the nice
picture of the string field gently oscillating around the vacuum with
decreasing amplitude, takes a serious hit. Indeed, we show that
already at level two, the tensor fields which must be included in our
analysis, bring derivatives into the equations of motion in such a way
that these become a so called \textit{system of higher order neutral
  DDEs} with one positive delay.  Such DDEs cannot in general be
solved with the method of steps. We have to do some simplifications
before obtaining numerical solutions. What we can do, however, is to
look at the equations of motion close to the vacuum. We will find that
the latter effectively contain several delays.  This would pose no
further conceptual difficulty if all delays were positive, but we show
that we obtain {\em negative delays} as well.  In other words, the
equations of motion effectively involve the fields at some past
light-cone times, but also at some {\em future} light-cone times.  We
show that if we have both negative and positive delays, there exist
{\em growing oscillation} modes around the vacuum. This suggests that
the string field may not converge to the non-perturbative vacuum. This
seems to be in contradiction with the analytic solution obtained in
\cite{Hel-Sch}, and also with our numerical solution obtained by
expanding the fields in exponential series (this provides in principle
a very accurate solution but only up to a limited light-cone
time). But the two pictures can be reconciled if we notice that the
initial conditions for the analytic solution (which are the same as
those for the exponential series solution) are very special. The
diverging modes might not be excited for this particular solution, but
our results imply that a small change in the initial conditions can
render the rolling non-convergent.

Thirdly, by studying the equations of motion near the vacuum at level
four, we show that there are more delays at this levels, and that they
are more spread, both towards the past and towards the future. We are
led to conjecture that in the large level limit, one recovers a
discrete version of the non-locality (\ref{tilde}). Subsequently
setting the dilaton gradient to zero, the delays become less and less
spaced, and we will recover (\ref{tilde}). We conclude that in
particular, the nice simplification of the non-locality that happens
at level zero, is only accidental.

This paper is structured as follows: In the next section, we calculate
the action and derive the equations of motion at level two. We give a
short review of DDEs in Appendix \ref{s_dde} and the complete results
are given in Appendices \ref{Lagrangian24} and \ref{EoMs24}. We solve
the level-zero equation with the method of steps, and we attempt to do
the same at level two. We show what problems we face and what can be
done to get some information on the rolling at this level. In Section
\ref{s_oscillations}, we study the linearized equations of motion near
the vacuum. We show that negative delays appear at level two and four
and conclude the general form of non-locality in the linear dilaton
background. In order to do so, we need to calculate the determinants
of polynomial matrices. A naive approach fails if the matrices are too
large, so we explain a little-known method for calculating such
determinants in Appendix \ref{polydet_s}. In Section
\ref{s_discussion}, we discuss further the consequences of our
results. And at last, we offer a review of linear dilaton CFT in
Appendix \ref{rev lin Dil}.

%%%%%%%%%%%%%%%%%%%%%%%%%%%%%%%%%%%%%%%%%%%%%%%%%%%%%%%%%%%%%%%%%%%%%%%%%%%%%%
\section{Light-like dynamics of the vacuum transition in Siegel gauge}
\label{s_level2}
%%%%%%%%%%%%%%%%%%%%%%%%%%%%%%%%%%%%%%%%%%%%%%%%%%%%%%%%%%%%%%%%%%%%%%%%%%%%%%

This section is divided into several paragraphs. At first we show how to 
derive the equations of motion of open string field theory in a linear
dilaton background with level truncation. We present more details of the
calculation, on the one hand to introduce the notation, and on the other hand
because they are  omitted too often. After that we briefly describe how
we solved the resulting delay differential equations on the computer. 
To that end we explain how astonishingly natural it is in our setup to 
choose the infinitely many initial conditions required for producing a unique solution.
We test our machinery in the simplest possible case of level zero, and observe
excellent agreement with the literature \cite{Hel-Sch,Barnaby}. We then go beyond
level zero and explain our results at level two and four, which
can be summarized as follows:
The individual modes of the string field are initially in the perturbative vacuum,
then, driven by the tachyon, they grow steeply. Finally they oscillate around 
their respective vacuum expectation values with decaying amplitudes.

\paragraph{Derivation}

If we write the string field in terms of vertex operators as 
$|\Psi\rangle = \Psi(0) |0\rangle$, the action of Witten's open string field theory reads
\begin{equation}
S = - \frac{1}{g^2} \, \left( \frac{1}{2} \langle \Psi , Q  \Psi \rangle + 
\frac{1}{3} \langle f_1 \circ \Psi(0) \, f_2 \circ \Psi(0) \, f_3 \circ \Psi(0) \rangle 
\right),
\end{equation}
where the functions $f_i$ are the conformal transformations mapping
each string (semi-disk) to the common interaction upper-half plane. 

We have derived the action and equations of motion with two independent methods.
First we calculated the action by hand, calculating explicitly the
conformal transformations $f_i \circ \Psi(0)$, and then the
CFT correlators
\footnote{For a detailed derivation of the correlator in the 
linear dilaton background, cf. \cite{Ho:2007ar}.}
\footnote{Note that we use the complex derivative $\del_{z}$ instead of the real derivative.}
:
\begin{eqnarray}
\corr{\prod_{i=1}^{n}\ordB e^{ik_{i}\cdot X\left(z_{i}\right)}\ordB\prod_{j=1}^{p}\del_{z'_{j}}X^{\mu_{j}}\left(z'_{j}\right)} & = & \left(2\pi\right)^{D}\delta^{D}\left(\sum_{i}k_{i}\right)\prod_{i,,j=1,i<j}^{n}\left|z_{i}-z_{j}\right|^{2\alpha'k_{i}\cdot k_{j}}\nonumber \\
 &  & \cdot\corr{\prod_{j=1}^{p}\left[v^{\mu_{j}}\left(z'_{j}\right)+q^{\mu_{j}}\left(z'_{j}\right)\right]}.\label{eq:corr n exp. and delX fields}\end{eqnarray}
The new objects $v,q$ serve as a tool to quickly work out the combinatorics
of the contractions - just expand the product into a polynomial in
$v,q$ and observe the following rules:

\begin{enumerate}
\item \emph{Replace $v^{\mu}\left(z\right)=-i\alpha'\sum_{i=1}^{n}\frac{k_{i}^{\mu}}{z-z_{i}}.$
\label{rule v}}
\item \emph{Contract products of two $q$s using $-\alpha'\eta^{\mu\nu}/2\left(z-z'\right)^{-2}.$\label{rule q^2}}
\item \emph{Remove all terms with an odd number of $q$s. \label{rule odd q}}
\item Note that the general expression diverges if $z=z'$ . For correlators
of normal ordered products (e.g. $\ordB\del X^{\mu}\, e^{ik\cdot X}\ordB\times\ordB\dots\ordB$)
these terms precisely cancel, providing a non-singular result. In
this case we can further simplify with the additional rule\\
\emph{Remove all terms in $v$ with $z=z_{i}$ and all $q$-products
at the same point, $z=z'$. \label{rule remove at same point}}
\end{enumerate}
The dilaton background enters explicitly in two ways:
\begin{enumerate}
\item the delta function is updated to include the breaking of the translation
invariance by the linear dilaton background \[
\delta^{D}\left(\sum_{i}k_{i}\right)\to\delta^{D}\left(\sum_{i}k_{i}+iV\right),\]
where the delta function of a complex argument is formally defined
by the following integral representation\begin{equation}
\delta^{D}\left(\sum_{i}k_{i}+iV\right)\equiv\frac{1}{\left(2\pi\right)^{D}}\int\mbox{d}^{D}x\,\, e^{i\, x\cdot\sum k_{i}-V\cdot x}.\label{eq:delta function with dilaton}\end{equation}

\item through the modified conformal transformation law \[
X^{\mu}\left(z,\bar{z}\right)\to f\circ X^{\mu}\left(z,\bar{z}\right)=X^{\mu}\left(f\left(z\right),f\left(\bar{z}\right)\right)+\frac{\alpha'}{2}V^{\mu}\log\left|f'\left(z\right)\right|^{2}\]
 needed when mapping the string field vertices to the interaction worldsheet.
\end{enumerate}

In order to make sure that our results are correct, we redid the same
calculation with the method of conservation laws. Luckily, the
conservation laws for an anomalous vector (like $\partial X^\mu$ in a
linear dilaton background) were already worked out by Rastelli and
Zwiebach in \cite{Rastelli:2000iu}. This method has the advantage of
being easy to implement on a computer. In our case we wrote
a \texttt{mathematica} program. To our satisfaction both methods
agreed entirely. Since the calculations become rather cumbersome at
higher levels to do by hand, we relied on our code for the equations
of motion at levels (2,6) and (4,8).

Explaining our notation we will quickly see that at level two, the
string field written out in its mode expansion can be reduced to
just eight spacetime fields. We follow the conventions by \cite{Kos-Sam} 
(except that we call $\beta$ their $\beta_1$) and write
the string truncated to level two fields. Working in the Siegel gauge,
we can eliminate those terms containing a $c_{0}$ -ghost mode, and
due to the twist symmetry of the action, we can consistently
set  all terms at odd levels to zero. This leaves us with the following expression for the string field
at level two:

\begin{equation}
\state\Psi=\biggl\{\phi+\frac{i}{\sqrt{2}}B_{\mu}\alpha_{-2}^{\mu}+\frac{1}{\sqrt{2}}B_{\mu\nu}\alpha_{-1}^{\mu}\alpha_{-1}^{\nu}\,+\beta b_{-1}c_{-1}\biggr\} c_{1}\state0.\label{eq:SF at level two}\end{equation}

Since we work in $D=26$ spacetime dimensions, expr.  (\ref{eq:SF at level two})  contains
379 spacetime fields. But in the case of light-like tachyon rolling,
we can drastically reduce the number of fields needed in our calculation.
Working in the light-cone frame 
we can split the dimensions into light-like and ordinary components
\[
\mu=\left(0,1,2,\dots D-1\right)\to\left(+,-,2,3,\dots D-1\right)\equiv\left(+,-,i\right).\]

We assume the linear dilaton gradient light-like, $V^{2}=0$. By rotational
symmetry we can choose a coordinate system where $V=\left(V^{+},0,\dots,0\right)$.
Furthermore we consider spacetime fields $\phi,\beta,B_{\mu},B_{\mu\nu}$
in expr. (\ref{eq:SF at level two}) that depend only on the first lightcone
coordinate $x^{+}$. As detailed in appendix \ref{Lagrangian24}, we can then focus
on the following eight (of 379) fields to compute the action:
 \begin{equation}
\left\{ \phi,\, B^{+},\, B^{-},\, B^{++},\, B^{+-},\, B^{--},\, F,\,\beta \right\}. \label{eq:eight fields}\end{equation}
Note that $\phi$ is the tachyon field and $F$ is the scalar field
associated with the contribution of  $B^{ij},~ i,j=2\dots25$ to the
 trace of  $B^{\mu\nu}$ by 
\begin{equation}
\mathrm{Tr}\ B^{\mu\nu}\equiv-2B^{+-}+F. \label{eq:def F} 
\end{equation}
The resulting action is presented in appendix \ref{Lagrangian24},
both in Lorentz covariant form and explicitly using (\ref{eq:eight fields}).
To check its correctness, one can take the limit of vanishing dilaton
gradient, $V\to0$ and compare to the action found in \cite{Kos-Sam}. Both
expressions agree as desired. 

With the action computed, we can proceed to deriving the equations
of motion in notationally compact manner. For a Lagrangian ${\mathcal L}\left(\phi,\del\phi,\del^{2}\phi\dots\right)$ containing
arbitrary orders of field derivatives $\partial^{n}\phi$, the Euler-Lagrange
equation is

\[
0=\frac{\partial\mathcal{L}}{\partial\phi}-\del_{\mu_{1}}\frac{\partial\mathcal{L}}{\partial\left[\del_{\mu_{1}}\phi\right]}+\del_{\mu_{1}}\del_{\mu_{2}}\frac{\partial\mathcal{L}}{\partial\left[\del_{\mu_{1}}\del_{\mu_{2}}\phi\right]}-\dots\]
In this notation the derivatives are \emph{not} symmetrized, their
order matters:\[
\frac{\partial\left[\del_{\mu_{1}}\del_{\mu_{2}}\dots\del_{\mu_{k}}\phi\right]}{\partial\left[\del_{\nu_{1}}\del_{\nu_{2}}\dots\del_{\nu_{k}}\phi\right]}=\delta_{\mu_{1}}^{\nu_{1}}\delta_{\mu_{2}}^{\nu_{2}}\dots\delta_{\mu_{k}}^{\nu_{k}}.\]
This is just a matter of more convenient bookkeeping as we sum over
all combinations of indices. In a compact notation we can define the
differential operator $\mathcal{D}^{\phi}$ which returns the
equation of motion for the field $\phi$ when applied to the Lagrangian $\L$ depending
on $\phi$ and possibly other fields to any order in the field derivatives:
\begin{eqnarray*}
\mathcal{D}^{\phi} & \equiv & \sum_{k=0}^{\infty} \left(-1\right)^{k}\del_{\nu_{1}}
\del_{\nu_{2}}\dots\del_{\nu_{k}}\frac{\partial}{\partial\left[\del_{\nu_{1}}\del_{\nu_{2}}
\dots\del_{\nu_{k}}\phi\right]}\\
\Rightarrow & \mathcal{D}^{\phi}\L & \stackrel{!}{=}0.\end{eqnarray*}
Let's see how to apply this in a concrete example, take a generic
interaction term from $\mathcal{L}$, e.g. 
$\tilde{A}(x^{+})(\partial^{\mu_1} \ldots \partial^{\mu_l}
\tilde{B}(x^{+}))
(\partial_{\mu_1} \ldots \partial_{\mu_l}
\tilde{\phi}(x^{+}))e^{V^{+}x^{-}}$
and compute its contribution to the equation of motion. When deriving the equation of motion for
the tachyon field $\phi$ we apply $\mathcal{D}^{\phi}$. We
use $\Box=-2\del_{+}\del_{-}+\del_{i}\del^{i}=-2\del_{+}\del_{-}$
because $\L$ is independent of the $x_{i}$-coordinate.
Fields with a tilde are defined by\[
\tilde{\phi}\left(x^{+}\right)=K^{\alpha'\Box}\phi\left(x^{+}\right)=\sum_{n=0}^{\infty}\frac{1}{n!}\left(\alpha'\log K\right)^{n}\left(-2\del_{+}\del_{-}\right)^{n}\phi.\]
Now derive the equation of motion:
\begin{eqnarray*}
\mathcal{D}^{\phi}\left\{\partial_{\mu_{1}}\dots\partial_{\mu_{l}}\tilde{\phi}\right\} 
 & = & \sum_{k=0}^{\infty}\left(-1\right)^{k}\del_{\nu_{1}}\dots\del_{\nu_{k}}\frac{\partial\left[\partial_{\mu_{1}}\dots\partial_{\mu_{l}}\tilde{\phi}\right]}{\partial\left[\del_{\nu_{1}}\del_{\nu_{2}}\dots\del_{\nu_{k}}\phi\right]}\\
 & = & \sum_{k,n=0}^{\infty}\left(-1\right)^{k}\del_{\nu_{1}}\dots\del_{\nu_{k}}\frac{1}{n!}\left(-2\alpha'\log K\right)^{n}\delta_{\mu_{1}}^{\nu_{1}}\dots\delta_{\mu_{l}}^{\nu_{l}}\dots\delta_{+}^{\nu_{k-1}}\delta_{-}^{\nu_{k}}\delta_{2n+l,k}\\
 & = & \sum_{n=0}^{\infty}\left(-1\right)^{2n+l}\frac{1}{n!}\left(-2\alpha'\log K\right)^{n}\del_{\mu_{1}}\dots\del_{\mu_{l}}\left(\del_{+}\del_{-}\right)^{n}\\
 & = & \left(-1\right)^{l}\del_{\mu_{1}}\dots\del_{\mu_{l}}e^{-2\alpha'\log\left(K\right)\del_{+}\del_{-}}.\end{eqnarray*}
Applying $\mathcal{D}^{\phi}$ to the whole term, it becomes apparent that it acts essentially 
as a translation operator when $\del_{-} \to V^{+}$:
\begin{eqnarray*}
&& \mathcal{D}^{\phi}\left\{\tilde{A}(x^{+})
\left(\partial^{\mu_1} \ldots \partial^{\mu_l}\tilde{B}(x^{+})\right)
\left(\partial_{\mu_1} \ldots \partial_{\mu_l}\tilde{\phi}(x^{+})\right)
e^{V^{+}x^{-}}\right\} \\ 
& = & \left(-1\right)^{l}\del_{\mu_{1}}\dots\del_{\mu_{l}} 
e^{-2\alpha'\log\left(K\right)\del_{+}\del_{-}}\left\{\tilde{A}(x^{+})
\left(\partial^{\mu_1} \ldots \partial^{\mu_l}\tilde{B}(x^{+})\right) 
e^{V^{+}x^{-}} \right\} \\
 & = & \left(-1\right)^{l}\del_{\mu_{1}}\dots\del_{\mu_{l}} 
e^{-2\alpha'V^{+}\log\left(K\right)\del_{+}}\left\{ A(x^{+})\left(\partial^{\mu_1} \ldots \partial^{\mu_l}B(x^{+})\right)e^{V^{+}x^{-}}\right\} \\
 & = &\left(-1\right)^{l}\del_{\mu_{1}}\dots\del_{\mu_{l}} \left\{ A\left(x^{+}-2\alpha'V^{+}\log\left(K\right)\right)\left(\partial^{\mu_1} \ldots \partial^{\mu_l}B\left(x^{+}-2\alpha'V^{+}\log\left(K\right)\right)\right)e^{V^{+}x^{-}} \right\}\end{eqnarray*}
where finally the tilde was eliminated because 
$\Box A(x^+) = 0$.
The translation operator simply shifts the argument $x^{+}$ by $-2\alpha'V^{+}\log\left(K\right)$.
This is a generic feature in \textit{every} interaction term, we therefore
define the symbol \[
y^{+}\equiv x^{+}-2\alpha'V^{+}\log\left(K\right)\]
for the shifted point to abbreviate the notation. The reason for 
introducing $\mathcal{D}^{\phi}$ is that it simplifies the calculation
significantly. As an example, consider the \textit{chain rule}
\[\mathcal{D}^{\phi}\left\{ \tilde{A}\left(x^{+}\right){\tilde{\phi}}^{2}
\left(x^{+}\right)e^{V^{+}x^{-}}\right\} = 2 A\left(\yp\right) \phi\left(\yp\right)e^{V^{+}x^{-}}.\]
The set of eight equations of motion contains derivatives of up to
fourth order, each equation is quite lengthy with one notable
exception: the equation of motion for $B^{--}$ is short and can be
solved easily:\[
\mathcal{D}^{B^{--}}\L_{total}=B^{++}(x^{+})+{B^{++}}'(x^{+})+\frac{8B^{++}(y^{+})\phi(y^{+})}{\sqrt{3}}\stackrel{!}{=}0.\]
The equation is linear in $B^{++}$, hence we set $B^{++}\equiv0$ to
obtain a solution. As a matter of fact this still admits non-trivial
solutions for the other seven fields. This observation was also made
in \cite{Erler}, in the related context of OSFT using the lightcone
basis for the modes of the string
field. 
This is a peculiarity of level (2,4). Indeed, already at level (2,6)
one cannot consistently set $B^{++}$ to zero anymore.

Setting $B^{++}$ to zero at level (2,4) reduces
the length of the equations of motion by about one third and the total differential
order from 21 to 17. The resulting set of equations of motion is presented in appendix
\ref{EoMs24}. The seven equations contain a total of 144 terms. For
reference we list the seven remaining fields\begin{equation}
\left\{ \phi,\, B^{+-},\, B^{--},\, F,\,\beta,\, B^{+},\, B^{-},\right\}. \label{eq:seven fields}\end{equation}
Let $\Psi_{j},\, j=1\dots7$,
denote one of the above fields, e.g. $\Psi_{1}\left(\xp\right)=\phi\left(\xp\right)$
is the tachyon field. Then the generic form of an equation of motion is
\footnote{To simplify the notation, we set $\Vp = \alpha' \equiv 1$.}:

\begin{equation}
0=\Psi_{j}\left(\xp\right)+\left(-1\right)^{\delta_{j,1}}\del\Psi_{j}\left(\xp\right)+
\sum_{i,k=1}^{7}\sum_{n,m=0}^{3}a_{nm}^{ik}\del^{n}\Psi_{i}\left(\yp\right)\del^{m}\Psi_{k}\left(\yp\right).
\label{eq: generic EoMs}\end{equation}

The first two terms come from the kinetic term in the action. They
have the same sign except for the tachyon and are evaluated at position
$\xp$. All derivatives are understood with respect to $\xp$. All
terms in the sum arise from the cubic interaction. They are evaluated
at $\yp$ and contain derivatives of up to the third order. In fact
many of the real-valued coefficients $a_{nm}^{ik}$ are zero.

\paragraph{Solving the Equations of Motion}

Given the general structure of the equations of motion 
(\ref{eq: generic EoMs}),
let us now focus on solving them at level $\left(2,4\right)$. 
There are derivatives with respect to $x^{+}$ at both points: $\xp$ (max: 1st order)
and $\yp=\xp-\gamma$ (max: 3rd order). The total
differential order of the system of equations is 17. Because of the
non-locality (fields at $\xp$ and at $\yp$) these are
not ordinary differential equations (ODEs). It is the key observation
that, mathematically speaking, we have a system of coupled \emph{delay
differential equations}
of the neutral type (NDDE) with one constant delay $\gamma$. The
theory of delay differential equations (DDE) has been developed to
great extent in the 20th century as this type of differential equation
arises in a large array of disciplines: populations dynamics, machine
control theory, neutron diffusion, spreading of diseases, retarded
propagation in classical electrodynamics etc. We give a short introduction
to DDEs, highlighting the differences to ODEs, in appendix \ref{s_dde}.
Further useful references are \cite{Bellman-Cooke,Driver,Bellen}. 

In order to solve DDEs we use the \emph{method of steps}, explained
further in Appendix \ref{s_dde}. The basic idea is to reduce the problem
of computing the solution over a full interval to subintervals where
the DDE reduces to an ODE which is solved with standard methods. In
order to obtain a unique solution, it is however not sufficient to
give an initial condition at one point as in the ODE case: initial
conditions over a finite interval (of the length of the delay $\gamma$)
have to be specified. These conditions are called the \emph{initial
data}. They are essentially uniquely determined when we require that
the tachyon condensation starts in the perturbative vacuum. 

There are other methods. For instance Barnaby et al. \cite{Barnaby}
transformed the level-zero DDE into an equivalent diffusion-like local
partial differential equation problem with suitable boundary
conditions, solved that with standard codes and finally transformed
back to obtain the DDE solution. For the system of equations at level
(2,4), their method seems cumbersome to apply. In contrast, the method
of steps is powerful enough to easily extend to the case of several
unknowns. For more details on diffusion methods, we refer the reader to

We have seven coupled equations for seven fields with
derivatives up to the third order. The numerical stability and
convergence properties of the method of steps have been studied
carefully in the past decades, cf. \cite{Bellen} for an extensive
review of numerical methods. The problem can now be considered a
standard one, much like solving a system of ODEs numerically.  This
implies that the user has to carefully check the consistency of the
numerical solution, preferably by independent methods. Much of the
effort in this section is aimed in that direction.

On the computer we need dimensionless numbers. As of now we work in
units where $\alpha'\equiv 1$. For convenience we choose $V^{+}\equiv1$.

\paragraph{Choosing initial data}

Let us now see how to supply initial data for each field to obtain
a unique solution to the equations of motion. A priori we could choose any initial
data, but physical reasons nearly completely fix them.
Consider as an example the tachyon field $\phi$. In principle we
would like to obtain the solution $\phi\left(\xp\right)\forall\xp\in\mathbb{R}$
from the equations of motion. On the computer we can only obtain a solution on some
finite interval $\left[\xp_\text{min},\xp_\text{max}\right]$. We then need to fix
the tachyon on $[\xp_\text{min}-\gamma,\xp_\text{min}]$, with the delay $\gamma=2\log K$.
To find constraints recall that we are interested in the tachyon condensation
solution: the solution should initially be in the perturbative vacuum
(string field $\state\Psi=0$) and in the end arrive at the non-perturbative
vacuum. Hence at sufficiently small $\xp_\text{min}$ the absolute value
of the tachyon and all other six fields $\Psi_{j}$$\left(\xp\right)$
from \eqref{eq:seven fields} should be small\[
\left|\phi\left(\xp_\text{min}\right)\right|\ll1, \qquad 
\left|\Psi_{j}\left(\xp_\text{min}\right)\right|\ll1\,.\,\]
As we take $\xp_\text{min}\to-\infty$ we can neglect all interaction terms
$\sim\Psi_{i}\Psi_{j}$ in the equations of motion \eqref{eq: generic EoMs} and
receive dramatically simplified equations. In other words, we linearize
the equations of motion around zero, $\Psi_{j}=0+\delta\Psi_{j}$ , and neglect
all terms quadratic in the small quantity $\delta\Psi_{j}$. The
resulting equations are\begin{equation}
0=\delta\Psi_{j}\left(\xp\right)+\left(-1\right)^{\delta_{j,1}}
\del\delta\Psi_{j}\left(\xp\right).\label{eq:linearized Eom}\end{equation}
These are first order \emph{ordinary} differential equations 
which can be solved analytically, we just need
to fix the initial condition. Starting in the unstable vacuum means
all fields vanish at negative infinity, 
\[
\Psi_{j}\left(-\infty\right)=0,\quad\forall j.\]
The linearized tachyon equation of motion has the solution
\begin{align}
& \delta\phi\left(\xp\right)=a_{\phi}e^{\xp}\label{eq:tach initial data}\\
& 0 =\delta\phi\left(-\infty\right)\quad \Rightarrow \quad a_{\phi}\mbox{ arbitrary}.\nonumber \end{align}
The solution grows exponentially, reflecting the instability of the
perturbative vacuum. The initial condition does not fix the constant
$a_{\phi}$. This is the only free parameter in choosing the initial
data as we will see shortly. In short, there are just three cases, \[
a_{\phi}=\begin{cases}
\mbox{positive}\\
0\\
\mbox{negative}\end{cases}\]
that give qualitatively different solutions to the full equations of motion. From equation (\ref{eq:tach initial data}) $a_{\phi}$ can be chosen to have any real
value. If we set it to zero 
this corresponds to the static solution in the perturbative vacuum
that we are \emph{not} interested in. 

\begin{figure}[!ht]
\begin{center}
\input{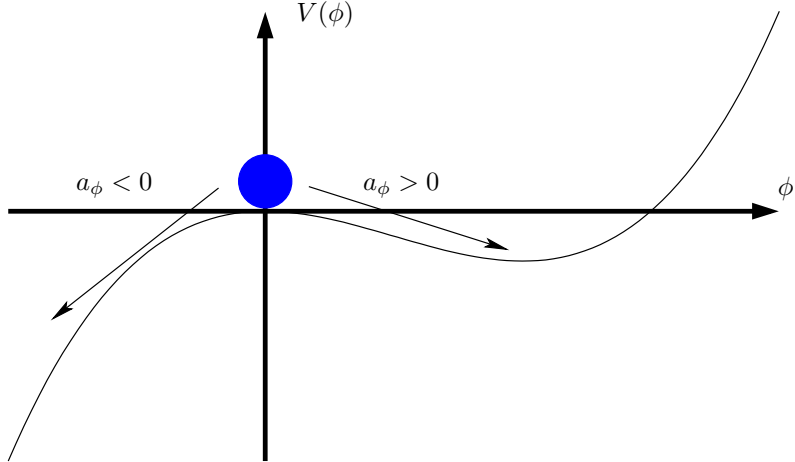}
\caption{\footnotesize{Effective tachyon potential. Depending on the sign
of the coefficient the tachyon either rolls off to infinity or to the non-perturbative
 vacuum.}}
\label{fig: EffTachyonPot}
\end{center}
\end{figure}

For $a_{\phi}\ne 0$ we obtain interesting solutions.
 By shifting the origin in the $\xp$
direction, we can fix the absolute value
\begin{equation}
|a_{\phi}|\equiv1\label{eq: 1st tachyon coefficient is one}\end{equation}
 without loss of generality. It turns out that for $a_{\phi}<0$
the solutions diverge. This can be intuitively explained as
``rolling down the wrong side of the hill'': the effective tachyon potential
is presented schematically in Fig.~\ref{fig: EffTachyonPot}.
The linearized equations of motion for the other fields have the solution
\begin{eqnarray*}
&&\delta\Psi_{j}\left(\xp\right)=a_{j}e^{-\xp}\\
&& 0 = \delta\Psi_{j}\left(-\infty\right)\quad \Rightarrow \quad a_{j}=0.\end{eqnarray*}
In conclusion the initial data are as follows: the tachyon rises exponentially
with a prefactor of choice, all other fields vanish. This confirms
that the tachyon drives the condensation process. The same in a formula
is\[
\delta\Psi_{j}\left(\xp\right)=\delta_{j,1}a_{\phi}e^{\xp},\qquad\xp_\text{min}-\gamma\le\xp\le\xp_\text{min}.\]
We require that the solutions of the equations of motion be analytic,
thus we can express the analytic initial data as  
\[\delta\Psi_{j}\left(\xp\right)=\sum c_{n}^{j} \cdot \left(\xp\right)^{n}.\]
With our choice for the tachyon $\delta\Psi_{1}=a_{\phi}e^{\xp}$,
we have fixed every coefficient, $c_{n}^{1}=\frac{a_{\phi}}{n!}$.
Conversely for the other fields we find $c_{n}^{j}=0$. 

We conclude: 
the solution of the linearized equation requires only one initial condition
($a_\phi$ for the tachyon), it then fixes the countably many initial conditions 
required for finding a unique solution of the full non-linear DDE.
Note however that in general a Lagrangian with derivatives of all orders
does \textit{not} require supplying countably initial conditions for
a unique solution, cf. \cite{Barnaby:2007ve} for a review of the 
initial value problem for linear equations. 
\paragraph{Programming details}
Several codes for NDDEs implementing the method of steps are available.
We choose to use \texttt{mathematica }in the version 7 as it
easily allows to further manipulate the equations symbolically besides
the capability of solving systems of NDDEs with the single command
\texttt{NDSolve}.%
\footnote{This feature was not available in previous versions.}
 \texttt{mathematica }also allows to do the numerics with arbitrary
precision, which we made use of, as the built in machine precision
was not quite satisfactory. 
We want to warn the reader that in the case of NDDEs with higher order
derivatives at delayed positions \texttt{mathematica} 
quickly returns results without any warning or error message.
However upon plugging the supposed solutions into the 
equations of motion we realized that they do {\em not} satisfy the
equations. As always when using numerical results, checking is crucial.
In those cases where \texttt{mathematica} yields correct results we
used the parameters and options listed in Table~\ref{NDSolve}.
\begin{table}
\begin{center}
\begin{tabular}{|c|c|c|c|}
\hline 
Method & AccuracyGoal & WorkingPrecision & MaxSteps\tabularnewline
\hline
\hline 
Adams & 15 & 20 & 50000\tabularnewline
\hline
\end{tabular}
\caption{\footnotesize{Standard options used with \texttt{mathematica 7's NDSolve }to numerically
solve the equations of motion\label{NDSolve}.}}
\end{center}
\end{table}

\paragraph{Warm up at level zero}

\begin{figure}
\begin{center}
\includegraphics[scale=0.8]{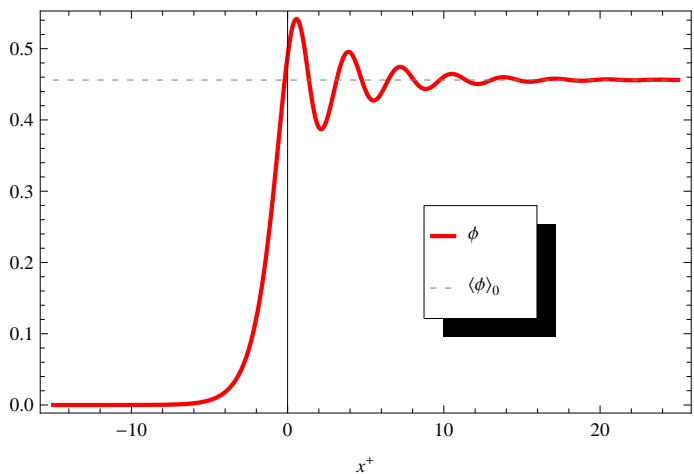}\caption{\footnotesize{
Level zero tachyon condensation calculated with the method of steps.
From the old vacuum $\phi=0$ we jump to the new vacuum with
exponentially dampened oscillations. The vev is indicated by the
dashed line.}}
\label{fig:level0}
\end{center}
\end{figure}

As a basic consistency check for our numerical method we run the
simplest example: the level zero truncation to the tachyon only. We
can compare the results to Barnaby et al. \cite{Barnaby} (diffusion
problem) and Hellerman and Schnabl \cite{Hel-Sch} (exponential series
solution) that each solved the same problem with a different method.
The equation of motion \cite{Hel-Sch} for the tachyon at level zero
is \begin{equation}
  0=\phi'\left(\xp\right)-\phi\left(\xp\right)+K^{3}\phi^{2}\left(\yp\right).\label{eq:hellerman
    EoM no alpha,V}\end{equation} We used the initial data
$\phi\left(\xp\right)=1\cdot e^{\xp}$, Eq.~\eqref{eq:tach initial
  data}, on an initial interval $[-25-2 \log K,-25]$.  In practice
this is close enough to the perturbative vacuum, as $
\phi\left(\xp\right)=e^{-25} \ll 1$. Moreover, the number of intervals
between $x^+=-25$ and $x^+=0$ is $\frac{25}{2 \log K} \approx 47.8$. This
tells us (see Appendix \ref{s_dde}) that the numerical solution will be
differentiable at least $47$ times for $x^+>0$; we can therefore
expect that it will be a very good approximation to the analytic
solution. The solution is depicted in Fig.~\ref{fig:level0}.
It is in excellent agreement with solutions
from \cite{Hel-Sch,Barnaby}. The asymptotic behavior
for large $\xp$ is known, it is of the form $exp\times cos$. The
frequency and decay rate can be determined by a simple fit, the values
agree perfectly with those noted by Hellerman/Schnabl from linearization
at large $\xp$. Note that these authors could compute the solution
of \eqref{eq:hellerman EoM no alpha,V} only as far as $\xp_\text{max}=7$
because of computing time limitations: the computational complexity
grows exponentially with $\xp$ for their method. Even though we plot
the solution only up to $\xp_\text{max}=25$ we compute it up to $\xp_\text{max}=100$
and beyond without any difficulty. The numerical solution ($\sim$0.1s)
actually takes less computing time than plotting the result ($\sim$1s)
on a modern computer. This confirms that the method of steps works
both fast and accurately.

We want to emphasize that the level zero equation of motion is in many facets simpler
than the level two equations of motion. Obviously it is only one equation compared to
seven coupled equations, but the major difference is of another kind:
the level two equations are \emph{neutral} DDEs, while \eqref{eq:hellerman EoM no alpha,V}
is of the {\em retarded} type, there are no derivatives of $\phi$ at position $\yp$.

\begin{figure}
\begin{center} 
\includegraphics[scale=1.1]{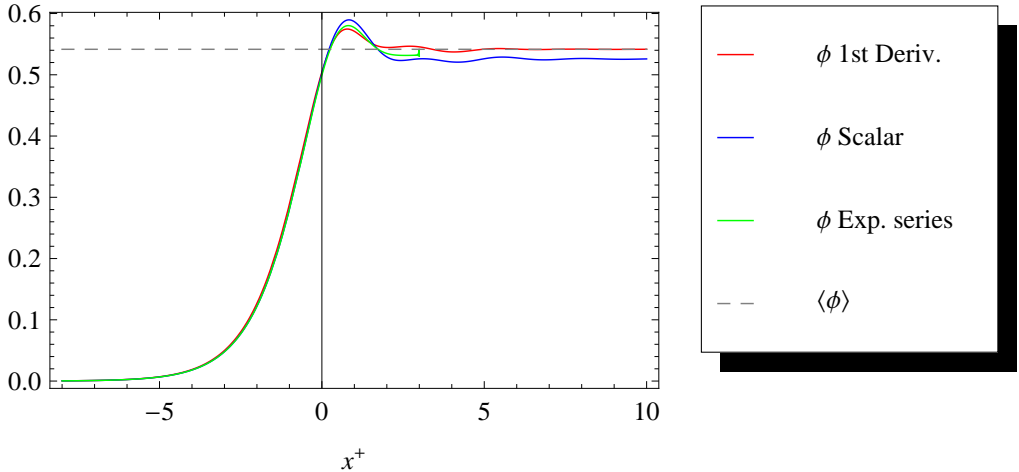}
\caption{\footnotesize{The different approximations for the tachyon
    condensation at level (2,4): From the perturbative vacuum $\phi=0$
    the tachyon jumps to the non-perturbative vacuum.  The vev of the
    full system is indicated by the dashed line. When ignoring higher
    derivatives the vev is unchanged. In contrast when considering the
    reduced system composed of only the scalar fields $\phi, F, \beta$
    the tachyon vev is slightly altered; this can be understood from
    the remark starting before Eq.~(\protect \ref{vacuumremark}).}}
\label{fig:CompTach}
\end{center}
\end{figure}

\paragraph{Level two solutions}
The full set of seven equations of motion at level (2,4) (appendix
\ref{EoMs24}) contains derivatives up to the third order in the
fields at the delayed position $y^{+}=x^{+}-\gamma$, and up to first
order at $\xp$, the total differential order is 17, while the total
number of terms is a staggering 144. 
As explained above  \texttt{mathematica} 7 cannot handle 
higher derivatives at delayed positions, and 
as of the time of writing we know 
of no other numerical method for solving a system of higher 
order neutral DDEs.
Hence we look for ways to simplify the problem that allow us to 
follow the vacuum transition:
\begin{enumerate}
\item Set all higher field derivatives to zero.
\item Consider only the scalar fields, $\phi,\beta,F$, and set the other 
vector/tensor component fields, 
$B^{+-},B^{--},B^{+},B^{-}$ to zero. Then the system of equations
contains at most first order derivatives.
\item Rewrite the fields as exponential series,
\begin{equation}
\Psi_j(x^+) = \sum_{n=1}^\infty a_{j,n} \, e^{n x^+},
\label{expseries}
\end{equation}
and solve for the first few hundred 
coefficients $a_{j,n}$ recursively. Evidently
for large $\xp$ this procedure requires knowing many of the $a_{j,n}$,
in fact the number of coefficients needed for an accurate solution 
grows exponentially with $\xp$. Thus we compute the solution with 
this approach in 
reasonable time only on a relatively small range $[\xp_\text{min}, 
\xp_\text{max}]$, with $\xp_\text{max}\approx 4.1$.
In this range the numerical solution is very accurate, but
for larger $\xp$, the last exponential in (\ref{expseries}) dominates and  
the numerical solution diverges. 
\end{enumerate}
Note that the range  $[\xp_\text{min}, \xp_\text{max}]$ is sufficient
to compare the different approximations and the different levels (0,0), (2,4), (2,6), (4,8) 
around the transition to the non-perturbative
vacuum, see Fig.~\ref{fig:CompTach} as an example for the tachyon only.
Similarly to the level zero solution, Fig.~\ref{fig:level0},
the tachyon is initially in the perturbative vacuum $\phi(\xp)=0$, 
then grows exponentially near $\xp = 0$, slightly overshoots the 
vacuum expectation value $\langle \phi \rangle_{(2,4)} $, then settles
in the non-perturbative vacuum. 
However using the exponential series ansatz we cannot evaluate the 
convergence properties around
the non-perturbative vacuum. We devote Section \ref{s_oscillations} to this issue.

All three methods agree very well up to the maximum value 
near $\xp \approx 1$, where the non-linear couplings due to the interaction 
between the various fields become important. But even for $\xp > 1$, 
the solutions are qualitatively very similar. 

\begin{figure}
\begin{center} 
\includegraphics[scale=1.1]{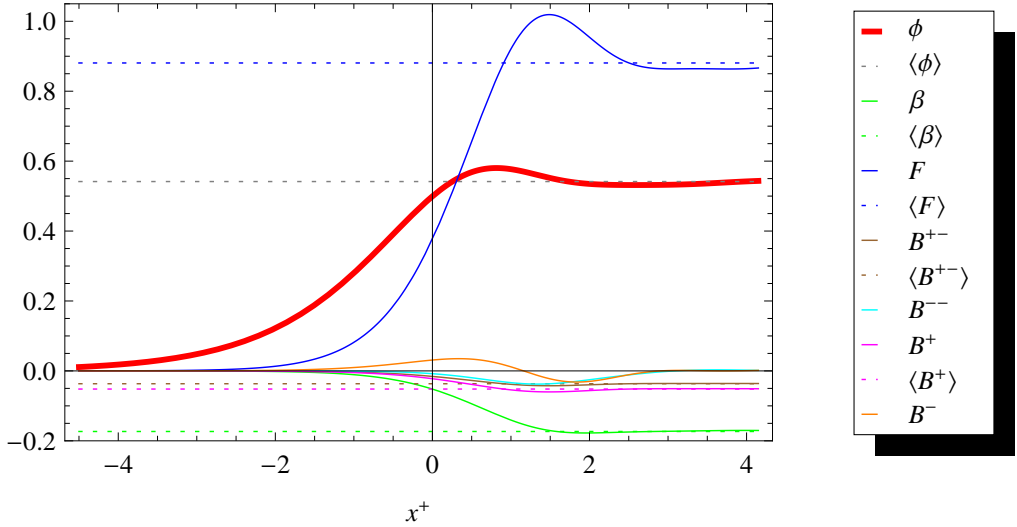}
\caption{\footnotesize{Level (2,4) tachyon condensation:
From the perturbative vacuum $\phi=0$ we jump to the new vacuum. The solution was calculated
using the exponential series ansatz with 500 terms and a precision of
150 digits.}}
\label{fig:ExpAll} 
\end{center}
\end{figure}

\begin{figure}
\begin{center} 
\includegraphics[scale=0.8]{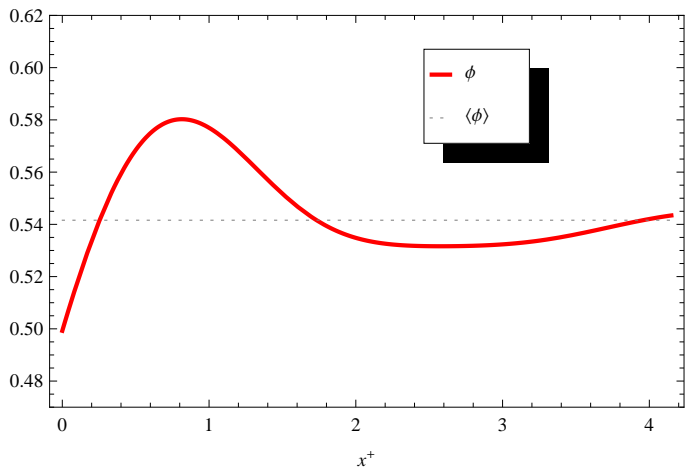}
\caption{\footnotesize{Level (2,4) tachyon condensation: Zoom on the
    tachyon $\phi(x^+)$ as it oscillates around the vev. The solution
    was calculated using the exponential series ansatz with 500 terms
    and a precision of 150 digits.}}
\label{fig:ExpTachyon} 
\end{center}
\end{figure}

Let us now consider the solution for the full system of the seven fields 
at level two that we calculated up to $\xp_\text{max} \approx 4.1$ using the exponential series ansatz 
(\ref{expseries}), 
see Figure \ref{fig:ExpAll}. 
The general picture involving the three 
stages 
$$ \mathrm{perturbative~vacuum} \to \mathrm{transition} 
\to \mathrm{non-perturbative~vacuum}$$
holds for all fields. The tachyon is drawn with a thick red brush 
to emphasize its importance for the vacuum transition: it is the first 
component to grow, and thus drives the others out of the perturbative 
vacuum. This affirms the intuitive notion of the tachyon as the unstable
mode of the string field, indicating the instability of the supporting
D-brane. 
In addition to the evolution of the seven fields, we included the vevs 
in the non-perturbative vacuum.
In Figure \ref{fig:ExpTachyon} we zoom on the tachyon in order to show how it
oscillates around the vev. 
Typically one would expect that all
fields, except the scalars, have vanishing vev, preserving translation
invariance in the non-perturbative vacuum.  But due to the presence of
the linear dilaton background $V\cdot x=V^{+}x^{-}$, translation
invariance is broken in the $\xp$-direction. Hence
e.g. $\corr{B^{+}}_{V^{+}=1}\ne0$ is no upset. If $V^{+}=0$, then
$B^{+}$ can have no vev.
How exactly the vevs are reached
for large $\xp$ is studied in detail in Section \ref{s_oscillations}.

There is more than one way to determine the vevs. The 
simplest one is the following: Simplify the equations of motion (appendix \ref{EoMs24}) to allow 
only constant solutions.
This results in a set of equations where the vevs $\corr{\phi},\corr{ B^{+-}},\corr F$
appear quadratically, all others linearly, hence there exist $3\cdot2+4=10$
solutions. With \texttt{mathematica} they are found in closed form,
as expressions depending on roots of 10th order polynomials. Numerically
the vevs can be evaluated to arbitrary precision. To pick the right
solution from the set of ten, it is sufficient to consider the tachyon
vev: 
\begin{itemize}
\item Two solutions can be eliminated as $\corr{\phi}\in\mathbb{C}$.
\item Six more can be neglected because $\corr{\phi}<0$. From the effective
potential, Fig.~\ref{fig: EffTachyonPot}, we know it is unbounded
for negative field values, hence there can be no finite negative vev.
\item One solution has $\corr{\phi}=0$. The vevs of the other fields vanish,
too. This is the unstable, \emph{perturbative vacuum} solution.
\item The last is the \emph{non-perturbative} \emph{vacuum} solution: $\corr{\phi}\approx0.5416$.
The resulting vevs to four significant digits are summarized in 
Table~\ref{tab:Vacuum-expectation-values:}.
\end{itemize}

\begin{table}[h]
\begin{center}
\begin{tabular}{|c|c|c|c|c|c|c|}
\hline 
$\corr{\phi}$ & $\corr F$  & $\corr{\beta}$ & $\corr{B^{+-}}$ & $\corr{B^{--}}$\
 & $\corr{B^{+}}$ & $\corr{B^{-}}$\tabularnewline
\hline
\hline 
0.5416 & 0.8808 & -0.1733 & -0.03670 & 0 & -0.05190 & 0\tabularnewline
\hline
\end{tabular}
\caption{\footnotesize{Vacuum expectation values at level two: the set of constant solutions of the full
equations of motion corresponding to the non-perturbative vacuum.}\label{tab:Vacuum-expectation-values:} }
\end{center}
\end{table}

Another way to determine the vevs is to realize that if we expand the
string field in the {\em universal} basis (i.e. using ghost modes and
matter Virasoro modes, but no matter oscillators), the vevs would be
independent of the dilaton gradient, and therefore equal to their
values in flat spacetime. Using the expression of the Virasoro
operators in terms of oscillators and dilaton gradient (\ref{eq:lin
  dil Virasoros}) we can then deduce the vevs of the fields in the
non-universal basis. Explicitly, at level two in the universal basis
and in Siegel gauge, we have three fields $t$, $u$ and $v$, and the
string field in the non-perturbative vacuum is given by
\begin{equation}
|\Psi \rangle = \corr{t} c_1 |0\rangle + \corr{u} c_{-1} |0\rangle + \corr{v} L^m_{-2} c_1 |0\rangle,
\label{vacuumremark}
\end{equation}
where the vevs $\corr{t}$, $\corr{u}$ and $\corr{v}$ are well known 
\cite{Sen:1999nx}. Plugging the expression (\ref{eq:lin dil Virasoros}) for $L^m_{-2}$, we have 
\begin{equation}
|\Psi \rangle = \corr{t} \, c_1 |0\rangle + \corr{u} \, c_{-1} |0\rangle + \frac{1}{2} \corr{v} \, \alpha_{-1}^i \alpha_{-1}^i c_1 |0\rangle - \corr{v} \, \alpha_{-1}^+ \alpha_{-1}^- c_1 |0\rangle + 
\frac{i}{\sqrt{2}} V^+ \corr{v} \, \alpha_{-2}^- |0\rangle.
\end{equation}
Comparing with our string field expansion Eq.~(\ref{eq:SF at level two}), and the definition of F (\ref{eq:def F}), we find
\begin{equation}
\corr{\phi} = \corr{t}, \quad  \corr{\beta} = -\corr{u}, \quad  \corr F = 12 \sqrt{2} \corr{v}, \quad 
\corr{B^{+-}} = -\frac{1}{\sqrt{2}} \corr{v}, \quad \corr{B^{+}} = - V^+ \corr{v}.
\label{universalvev}
\end{equation}
And these give precisely the same values as in
Table~\ref{tab:Vacuum-expectation-values:}, providing further evidence
that our action was correctly calculated.

A curious result worth mentioning is that the following relations 
\begin{equation}
B^{+-} = -\frac{1}{24} F, \qquad B^+ = - \frac{V^+}{12 \sqrt{2}} F,
\label{lev24relations}
\end{equation}
which, by Eq.~(\ref{universalvev}), should hold for the expectation
values, actually hold {\em for all} $x^+$ at level (2,4). This can be
roughly seen on Fig.~\ref{fig:ExpAll}. We have checked these relations
numerically beyond doubt (they hold with a precision of at least 100
digits), but haven't found any simple reason to explain them. They
must in fact be ``accidental'' because they do not hold anymore at
level (2,6).

\begin{figure}
\begin{center} 
\includegraphics[scale=1.0]{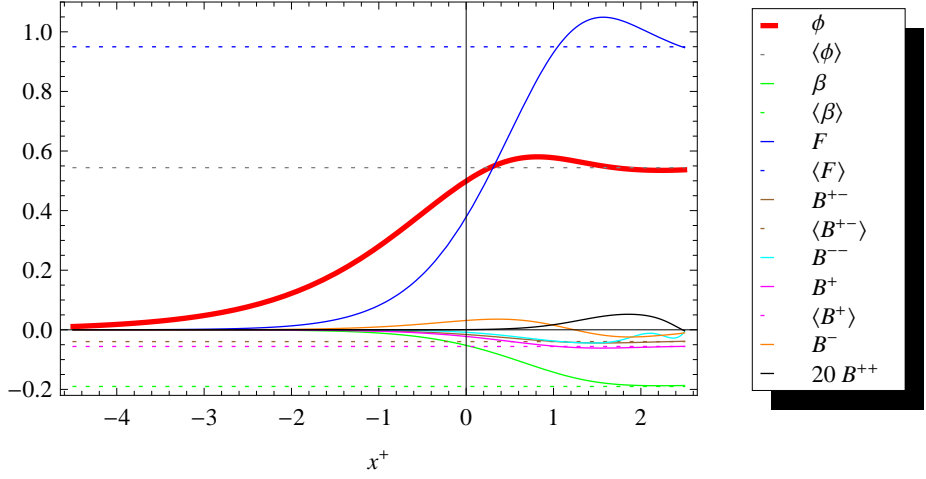}
\caption{\footnotesize{Level (2,6) tachyon condensation:
The solutions were calculated using the exponential series ansatz with
300 terms and a precision of 150 digits. We have multiplied $B^{++}$
by 20 in order to make it visible.}}
\label{fig:ExpAll26} 
\end{center}
\end{figure}

Let us now briefly look at level (2,6). The main difference here is
that we cannot set $B^{++}$ to zero anymore. In particular, the
equation of motion for $B^{--}$, which at level (2,4) contained only
terms proportional to $B^{++}$, now contains in particular a term
$(B^{--})^2$. We show the exponential series solution for the rolling
in Fig.~\ref{fig:ExpAll26}

%%%%%%%%%%%

%%%%%%%%%%%

\paragraph{Level four solutions}

\begin{figure}
\begin{center} 
\includegraphics[scale=1.1]{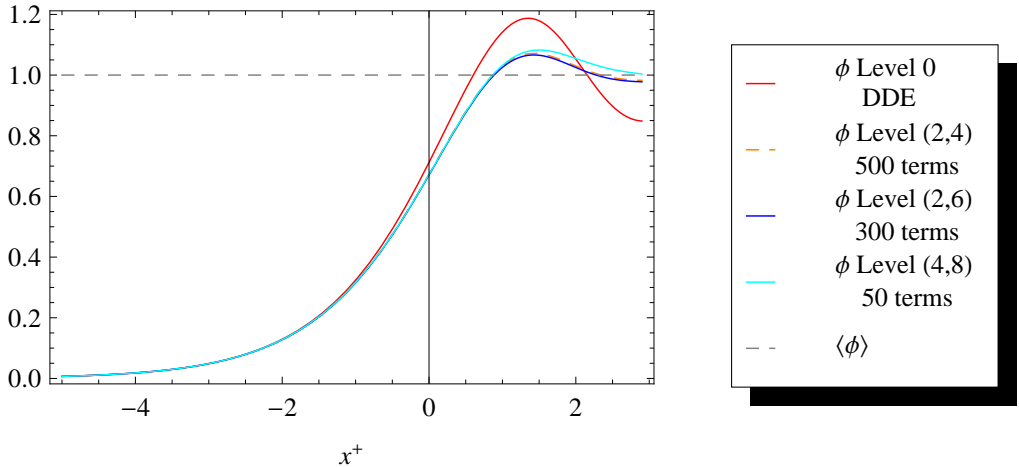}
\caption{\footnotesize{ Comparison of tachyon condensation at levels
    (0,0), (2,4), (2,6) and (4,8). The vevs have been normalized to
    unity. The solution at level zero was calculated with the method
    of steps, and the solutions at higher levels were calculated using
    the exponential series ansatz with the origin of $\xp$ chosen such
    that the first coefficient is one.  Accordingly the initial
    condition $a_{\phi}=K^{-3} $ was used for the level zero DDE.  The
    curves at levels (2,4) and (2,6) are almost indistinguishable.}}
\label{fig:lev2vs4} 
\end{center}
\end{figure}

Let us now turn to the system of equations at level (4,8). Based on the method of 
conservation laws we used our \texttt{mathematica} code to compute the 
equations of motion.
There are now 50 fields to consider, the equations of motion contain 21400 terms. 
We want to investigate to what extent
the solution changes compared to level two. Again we used the exponential series
ansatz and computed the first 50 coefficients for all 50 fields. This 
calculation took about 15 h on a fast computer. 
From those 50 fields we decide to focus on the tachyon, discussion of
the other 49 fields would be redundant, for they are all qualitatively
similar. Now we compare the tachyon at level two and four, Figure
\ref{fig:lev2vs4}. 
As before the vevs of all fields are extracted
from the constant solutions. For the tachyon, the vev is only slightly 
bigger at level 4, the numerical value agrees with the estimate in \cite{Sen:1999nx}.
We notice that the tachyon overshoots more at level four, both
on a relative scale and in absolute value. 
As noted in \cite{Hel-Sch}, the full solution (no level truncation)
converges monotonically. One might have expected to see the
overshooting decrease monotonically with the level as well, since it is
a lot more pronounced at level zero than at level two, but that is not
the case. For completeness, the relative overshooting of the tachyon
is shown in Table~\ref{tab:overshooting}.
\begin{table}[h]
\begin{center}
\begin{tabular}{|c||c|c|c|c|}
\hline 
Level & 0  & (2,4) & (2,6) & (4,8)\tabularnewline
\hline
\hline 
$\left(\phi_\text{max}-\corr\phi\right)/\corr\phi$ & 18.7\%  & 7.1\%  & 6.7\% & 8.2\% \tabularnewline
\hline
\end{tabular}
\caption{{\footnotesize {Relative overshooting of the maximum vs the vev for
the tachyon at the lowest levels; it is not monotonically decreasing.}}}
\label{tab:overshooting} 
\end{center}
\end{table}

%%%%%%%%%%%%%%%%%%%%%%%%%%%%%%%%%%%%%%%%%%%%%%%%%%%%%%%%%%%%%%%%%%%%%%%%%%%%%%
\section{Oscillations around the non-perturbative vacuum}
\label{s_oscillations}
%%%%%%%%%%%%%%%%%%%%%%%%%%%%%%%%%%%%%%%%%%%%%%%%%%%%%%%%%%%%%%%%%%%%%%%%%%%%%%

In this section, we want to consider the equations of motions when the
fields are very close to their non-perturbative vacuum expectation
values. This analysis can be started in a relatively general
formalism, independent of the level. We start by putting the
components of the string field into a column vector
\begin{equation}
{\mathbf \Psi}(x^+) = \left(\phi(x^+), B^+(x^+), B^-(x^+), B^{+-}(x^+), 
\ldots \right)^\mathrm{T}.
\end{equation}
The length $n$ of this vector will be seven at level (2,4), eight at
level (2,6), and fifty 
at level (4,8), the highest level that we consider in this paper.
Next we write
\begin{equation}
{\mathbf \Psi}(x^+) = {\mathbf \Psi}_0 + \delta{\mathbf \Psi}(x^+),
\end{equation}
where ${\mathbf \Psi}_0$ is the non-perturbative vacuum expectation
value of ${\mathbf \Psi}$, and $\delta{\mathbf \Psi}(x^+)$ is a
small perturbation. We now write schematically the equations of motion
for $\delta{\mathbf \Psi}(x^+)$ at linearized level, i.e. keeping
only the terms that are linear in $\delta{\mathbf \Psi}(x^+)$. Note
that because ${\mathbf \Psi}_0$ is a solution of the full equations of
motion, there will be obviously no constant term in the linearized
equations. The general structure of these equations is easy to
understand. On the left hand side, we will write the contributions
from the kinetic term; this involves the fields at light-cone time
$x^+$ (not retarded) and at most first derivatives of the fields. On
the right-hand side we will write the contributions from the
interaction term. These all involve the same retarded light-cone time
$x^+-\gamma$, and at most $3L$ derivatives, where $L$ is the level.
 This is so because the
number of derivatives is at most equal to the total number of Lorentz
indices carried by the interacting fields, and each field of level $L$
can carry at most $L$ indices. So the general form of the equations of
motion is
\begin{align}
  & \delta{\mathbf \Psi}'(x^+) + A \, \delta{\mathbf \Psi}(x^+) = \nonumber \\
  & B_0 \, \delta{\mathbf \Psi}(x^+-\gamma) + B_1 \, 
\delta{\mathbf \Psi}'(x^+-\gamma) + 
  B_2 \, \delta{\mathbf \Psi}''(x^+-\gamma) + \ldots + 
B_{3L} \, \delta{\mathbf \Psi}^{(3L)}(x^+-\gamma),
\label{linearEoM}
\end{align}
where all the details are hidden in the $n$ by $n$ matrices $A$ and
$B_m$, $m=0,\ldots 3L$. We can now make the ansatz 
\begin{equation}
\delta{\mathbf \Psi}(x^+) = e^{\omega x^+} \, {\mathbf \Xi},
\label{oscillation_ansatz}
\end{equation}
where $\omega$ is a complex number and ${\mathbf \Xi}$ is a vector of
complex numbers. Since the equations (\ref{linearEoM}) are real and
linear, we can always find a real solution by adding its complex
conjugate to (\ref{oscillation_ansatz}). Note that with this ansatz,
all fields oscillate with the same frequency $\Im \omega$ with
exponentially decaying or growing amplitudes, according to the sign of
$\Re \omega$. The relative amplitudes and phase shifts between the
fields are encoded in ${\mathbf \Xi}$. Plugging this ansatz into
Eq.~(\ref{linearEoM}) and multiplying by $e^{\gamma \omega}$, we
obtain the equation
\begin{equation}
\left( \omega e^{\gamma \omega} \, I + e^{\gamma \omega} \, A - 
B_0 - \omega \, B_1 - \ldots - \omega^{3L} \, B_{3L} \right) \,{\mathbf \Xi} = 0, 
\label{deteq1} 
\end{equation}
where $I$ is the $n$ by $n$ identity matrix. After naming 
\begin{equation}
M(\omega) \equiv \omega e^{\gamma \omega} \, I + e^{\gamma \omega} \, A - 
B_0 - \omega \, B_1 - \ldots - \omega^{3L} \, B_{3L},
\label{M_def}
\end{equation}
we see that Eq.~(\ref{deteq1}) has a nontrivial solution for ${\mathbf \Xi}$ if and 
only if $\omega$ is such that
\begin{equation}
\det M(\omega) = 0.
\label{detM0}
\end{equation}
It is easy to see from (\ref{M_def}), that $\det M(\omega)$ is an {\em
  exponential polynomial} in $\omega$, i.e. a function of the form
\begin{equation}
\det M(\omega) = \sum_{j=0}^n p_j(\omega) \, e^{\beta_j \omega},
\qquad 0 = \beta_0 < \beta_1 < \ldots < \beta_n,
\label{exp_pol}
\end{equation}
where $p_j(\omega)$ are polynomials of degree $d_j$. In our particular
case, we have $\beta_j = j \gamma$. 

\begin{figure}
\begin{center}
\input{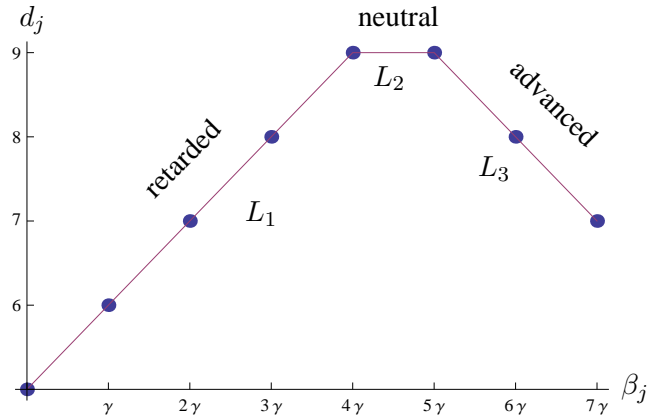}
\caption{\footnotesize{The distribution diagram of $\det M(\omega)$ at level (2,4). 
It has three segments, $L_1$ (retarded), $L_2$ (neutral), and $L_3$ (advanced) with 
respective slopes $\frac{1}{\gamma}$, $0$, and $-\frac{1}{\gamma}$.}}
\label{det24_f}
\end{center}
\end{figure}

\begin{figure}
\begin{center}
\input{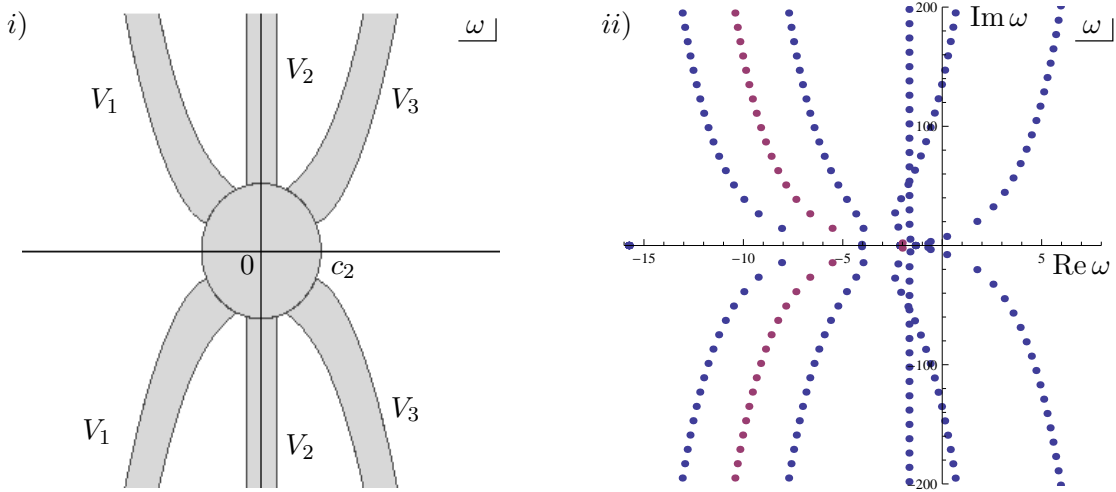}
\caption{\footnotesize{Location of the zeros of $\det M$ at level
    (2,4). $i)$: The asymptotic analysis tells us that there exists a
    $c_2>0$ such that outside the circle of radius $c_2$, all the
    zeros are located inside the retarded strip $V_1$, the neutral
    strip $V_2$ and the advanced strip $V_3$.}
  $ii)$: Roots found numerically for $-200<\Im \omega<200$. The blue
  dots are simple roots while the red dots are double roots.}
\label{zeros24_f}
\end{center}
\end{figure}

\paragraph{}
So we are interested in finding the roots of an exponential polynomial
(\ref{detM0}), and particularly in the signs of their real parts
because they will determine whether the perturbation $\delta{\mathbf
  \Psi}(x^+)$ will eventually die out or not. This problem has been
studied quite extensively (see for example \cite{Bellman-Cooke}). The
general answer is that one can write expressions for the asymptotic
values of the zeros (i.e. the zeros $\omega$ with $|\omega|
\rightarrow \infty$). More simply, one can use the {\em distribution
  diagram} to approximately locate the zeros. Namely, we plot the
points $P_j$ with coordinates $(\beta_j, d_j)$; we then draw the
convex polygonal line $L$ that joins $P_0$ to $P_n$, such that its
vertices are points of the set $P_j$ and such that no point $P_j$ lies
above it. This is illustrated in Fig.~\ref{det24_f} for the case of
level (2,4).
Now let us denote the successive segments of $L$ (from left to right) 
by $L_1, L_2, \ldots L_k$, and let $\mu_1, \mu_2, \ldots \mu_k$ denote
their slopes. Now the general result \cite{Bellman-Cooke} is that
there exist positive numbers $c_1$ and $c_2$ such that all the zeros
with norm greater than $c_2$, are located inside the union of the
strips $V_r$ defined by
\begin{equation}
\left| \Re (\omega + \mu_r \, \log \omega) \right| \leq c_1.
\end{equation}
It is clear that, for large $|\omega|$, these strips are located in
the left half-plane if $\mu_r$ is positive and in the right-half plane
if $\mu_r$ is negative. If $\mu_r=0$, the strip is vertical and
contains the imaginary axis. At level (2,4), we have three strips with
$\mu_r$ respectively equal to $\frac{1}{\gamma}$, $0$, and
$-\frac{1}{\gamma}$. These strips are sketched in
Fig.~\ref{zeros24_f} $i)$.
One can say a little more. In particular, in any region $R$ defined by
\begin{equation}
|\Re(\omega+\mu_r \log \omega)| \leq c_1, \quad 
|\Im(\omega + \mu_r \log \omega) - a| \leq b,
\label{ab}
\end{equation}
with no zero on the boundary, the number $n(R)$ of zeros in $R$ satisfies
\begin{equation}
1-n_r + \frac{b}{\pi} \beta \leq n(R) \leq \frac{b}{\pi} \beta + n_r - 1,
\label{zerodensity}
\end{equation}
where $n_r$ is the number of points of the distribution diagram on
$L_r$ and $\beta$ is the difference in values of $\beta_j$ between the
end-points of $L_r$. In Eq.~(\ref{ab}), $a$ is an arbitrary real
number, and $b$ is an arbitrary real positive number. So, for example
with $\mu_r =0$ (neutral strip), Eq.~(\ref{ab}) defines a rectangular
box of width $2 c_1$ and height $2 b$ centered on $a \times i$.
Therefore, Eq.~(\ref{zerodensity}) just means that the average
vertical density of the zeros along the neutral strip is given by
$\beta/\pi$. For the other branches with $\mu_r \neq 0$, the picture
is almost the same, just a little twisted. But in the limit $|\omega|
\rightarrow \infty$, $\log(\omega)$ is going to be roughly constant in
the box, so (\ref{ab}) defines an almost rectangular box centered on
$-\mu_r \log(\omega) + a \times i$. This means in particular that each
strip contains infinitely many zeros with arbitrarily large norm.  And
it has the following immediate consequence: {\em At level (2,4), $\det
  M(\omega)$ has infinitely many zeros with arbitrarily large positive
  real part.} This means that if we specify a generic initial
condition for the string field ${\bf \Psi}(x^+)$, $0 \leq x^+ \leq
\gamma$, with ${\bf \Psi}(x^+)$ close to the non-perturbative vacuum,
then we should expect that it will start to oscillate with
exponentially growing amplitude, at least until the linear
approximation becomes invalid.  For concreteness we show, in Figure
\ref{zeros24_f} $ii)$, the zeros of $\det M(\omega)$ found numerically
in the range $-200 < \Im \omega < 200$. The picture agrees with the
asymptotic analysis. We see in particular that the zeros are aligned
along branches in each of the strips, and that along these branches
the spacing between roots is asymptotically constant. Interestingly,
in the retarded strip there is a branch of {\em double roots} (the
double zeros are plotted in red in Figure \ref{zeros24_f} $ii)$). It
turns out that this is related to the fact, already mentioned in
Section \ref{s_level2}, that at level (2,4) there are only five
independent fields because of the relations (\ref{lev24relations}).

\begin{figure}
\begin{center}
\input{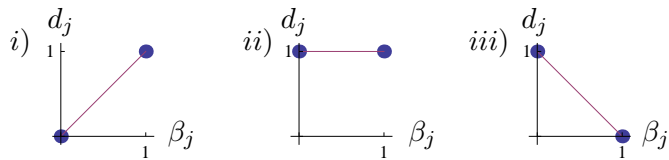}
\caption{\footnotesize{Distribution diagrams of some simple linear
    delay differential equations of $i)$ retarded type, $ii)$ neutral
    type, and $iii)$ advanced type.}}
\label{branches_f}
\end{center}
\end{figure}

In Fig.~\ref{det24_f} we have denoted the segment with positive slope
as {\em retarded}, the one with zero slope as {\em neutral} and the
one with negative slope as {\em advanced}. This denomination is easily
understood by looking at the following three simple linear delay
differential equations (we refer the reader to Appendix \ref{s_dde}
for definitions and more details on delay differential equations).
First, suppose we have an equation of the form
\begin{equation}
f'(t) = f(t-1).
\label{simple_retarded}
\end{equation}
This is an equation of the retarded type since the right-hand-side
includes $f$ at the retarded time $t-1$. Making the ansatz $f(t) =
e^{\omega t}$, we obtain the characteristic equation for $\omega$
\begin{equation}
\omega e^\omega - 1 = 0,
\end{equation}
whose distribution diagram is shown in Fig.~\ref{branches_f}$i)$. It
has one segment of positive unit slope; this justifies the
denomination ``retarded'' for such segments. 
Now we consider the neutral delay differential equation 
\begin{equation}
f'(t) = f'(t-1),
\label{simple_neutral} 
\end{equation}
which has for characteristic equation $\omega e^\omega -
\omega = 0$. Its distribution diagram, shown in
Fig.~\ref{branches_f}$ii)$, has one segment of zero slope, hence the
denomination ``neutral'' for such segments. At last, the equation 
\begin{equation}
f'(t) = f(t+1)
\label{simple_advanced}
\end{equation}
is of the {\em advanced} type because the right-hand side involves $f$
at the advanced time $t+1$. Its characteristic equation is $\omega -
e^\omega = 0$, whose branch diagram is shown on
Fig.~\ref{branches_f}$iii)$. It has one segment of negative slope,
which we therefore name $advanced$.

\begin{figure}[!ht]
\begin{center}
\input{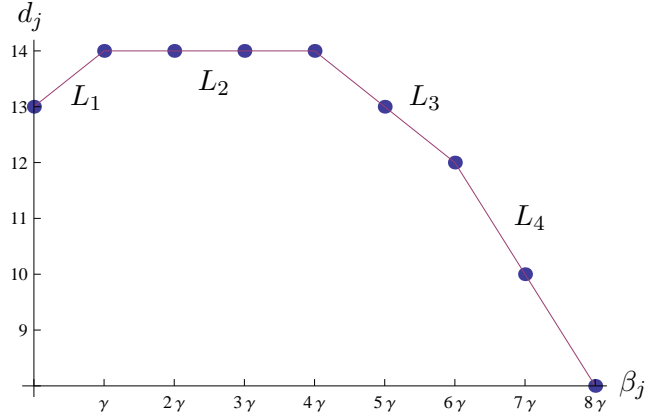}
\caption{\footnotesize{The distribution diagram of $\det M(\omega)$ at level (2,6). It has
four segments $L_1,\ldots,L_4$ with respective slopes $\frac{1}{\gamma}$, $0$, $-\frac{1}{\gamma}$, and $-\frac{2}{\gamma}$.}}
\label{det26_f}
\end{center}
\end{figure}

\begin{figure}[!ht]
\begin{center}
\input{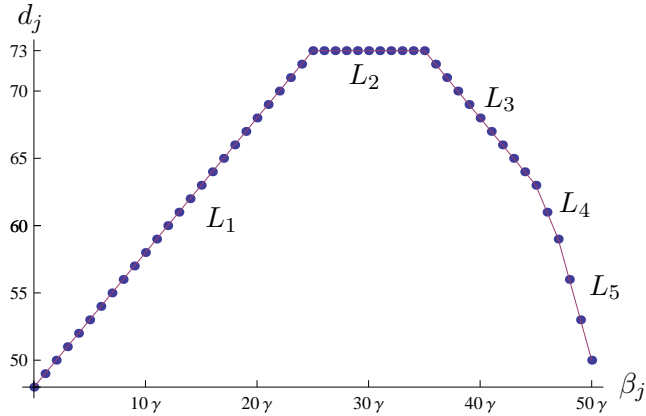}
\caption{\footnotesize{The distribution diagram of $\det M(\omega)$ at level (4,8). It has
five segments $L_1,\ldots,L_5$ with respective slopes $\frac{1}{\gamma}$, $0$, $-\frac{1}{\gamma}$, $-\frac{2}{\gamma}$, and $-\frac{3}{\gamma}$.}}
\label{det48_f}
\end{center}
\end{figure}

To summarize, at level zero the linearized equations of motion are
purely of the retarded type because the characteristic equation for
the tachyon is \cite{Hel-Sch}:
\begin{equation}
(\omega-1) e^{\gamma \omega} + 2 = 0,
\end{equation}
and its distribution diagram is thus as shown in
Fig.~\ref{branches_f}$i)$. From the above discussion, we therefore
know that its roots of large absolute value all have negative real
part, and it could thus have at most finitely many roots with positive
real part. It turns out that {\em all} the roots have negative real
parts, and the motion around the non-perturbative vacuum is stable at
this level. However, we have shown that at level (2,4) we obtain two
new segments, a neutral one and an advanced one which ruins the
stability around the non-perturbative vacuum. It is interesting to ask
what happens at higher levels. We show, in Fig.~\ref{det26_f}, the
distribution diagram found at level (2,6). 
Note that at this level, it is not anymore consistent to set
$B^{++}(x^+)$ to zero, so we have a total of eight fields. One of the
most striking differences with level (2,4) is that there is one more
advanced segment. This can be understood from the fact that the action
at level (2,6) contains higher derivatives at retarded times. One
notices also that the retarded segment $L_1$ has become much shorter.
With our code, we were able to calculate the action and equations of
motion up to level (4,8). From this large calculation we show in
Fig.~\ref{det48_f} the result for the distribution diagram.
At this level we have fifty fields. It turns out that the calculation
of $\det M(\omega)$ is a difficult problem. With the definition $z
\equiv e^{\gamma \omega}$, we see that the entries of the matrix
$M(\omega)$ are polynomials in the two variables $\omega$ and $z$.  We
explain, in Appendix~\ref{polydet_s}, why the computation of the
determinant of a polynomial matrix is hard and how to overcome this
difficulty. 

Interestingly, we see that at level (4,8), we have one retarded
branch, one neutral branch, and {\em three} advanced branches. In this
sense, the equations of motion become more and more advanced as the
level is increased. We can actually express this idea differently; one
could ask if we can write an {\em effective} equation of motion for
one field, e.g. the tachyon $\phi(x^+)$, that would give rise to the
same characteristic equation as Eq.~(\ref{detM0}). The answer is very
easy; indeed if we make the ansatz $\phi(x^+) = e^{\omega x^+}$, it is
immediately clear that the equation of motion
\begin{equation}
\sum_{j=0}^n p_j(\partial_+) \, \phi(x^+ + j\, \gamma) = 0
\end{equation}
is precisely the same as Eq.~(\ref{detM0}), where the $p_j$'s are the
polynomials defined by Eq.~(\ref{exp_pol}). We now do the usual
splitting of this differential equation by putting one of the terms
with highest derivatives on the left-hand side and all the other ones
on the right-hand side. So we choose a $j_{\text{max}}$ such that the
degree $d_{j_{\text{max}}}$ of the polynomial $p_{j_{\text{max}}}$ is
maximum (i.e. $d_{j_{\text{max}}} \geq d_j$, $j=0,\ldots n$). If the
distribution diagram has a neutral segment, then we can choose, for
$j_{\text{max}}$, any $j$ along this segment. After subtracting
$j_{\text{max}} \gamma$ from $x^+$, the equation of motion thus takes the form
\begin{equation}
p_{j_{\text{max}}}(\partial_+) \, \phi(x^+) = - \sum_{j \neq j_{\text{max}}} p_j(\partial_+)
\, \phi(x^+ + (j -j_{\text{max}}) \, \gamma).
\label{effdde}
\end{equation}
What we can immediately read from this equation is that, at level
higher than zero, the non-locality is not anymore concentrated at one
point in the past $x^+-\gamma$, but instead takes all the values $(j
-j_{\text{max}}) \, \gamma$, $j=0,\ldots n$. While this precise range
depends on the value of $j_{\text{max}}$ that we chose, it is clear
that it will extend in the future by, at least, an amount
$n_{\text{future}} \, \gamma$ equal to the length of the projection of
the advanced segments on the $\beta_j$ axis. And similarly, it will
extend in the past by, at least, an amount $n_{\text{past}} \, \gamma$
equal to the length of the projection of the retarded segments on the
$\beta_j$ axis. 

If we specify generic initial data for DDEs (\ref{effdde}) whose
lowest derivative appears at an advanced time (which we call advanced
DDEs), we should not expect that smoothing will occur when solving
with the method of steps. In fact, we show in Appendix \ref{s_dde} an
example of an advanced DDE which does not even possess a continuous
solution for a given initial data. For a system of advanced DDEs, on
the other hand, smoothing {\em may} occur. For our rolling solution,
we require an analytic solution. This is not in conflict with the
preceding remark because we are not free to choose any initial data,
as we have shown that the initial conditions are essentially uniquely
determined by demanding that the tachyon lives in the perturbative
vacuum in the infinite light-cone past.

It is clear from the distribution diagrams in Figs.~\ref{det24_f} and
\ref{det48_f}, that $n_{\text{future}}$ and $n_{\text{past}}$ both
increase from level (2,4) to level (4,8). Going from level (2,4) to
level (2,6), on the other hand, we see that although
$n_{\text{future}}$ increases, $n_{\text{past}}$ decreases. But if we
look only at levels $(L,2L)$, we nevertheless still expect that both
$n_{\text{future}}$ and $n_{\text{past}}$ will continue to grow beyond
level (4,8), and that in the limit $L \rightarrow \infty$, the
non-locality extends over the whole discrete range $\left\{n \,
\gamma, \ n \in \mathbb{Z}\right\}$.

At last, we remark that at infinite level, if we subsequently take the
limit $V^+ \rightarrow 0$, which is equivalent to 
shrinking the delay $\gamma \rightarrow
0$, the range of non-locality becomes continuous and extends over
{\em all} light-cone times. We thus recover, in this limit, the
 non-locality characteristic of usual (zero dilaton) string field
theory.

%%%%%%%%%%%%%%%%%%%%%%%%%%%%%%%%%%%%%%%%%%%%%%%%%%%%%%%%%%%%%%%
\section{Discussion and Conclusions}
\label{s_discussion}
%%%%%%%%%%%%%%%%%%%%%%%%%%%%%%%%%%%%%%%%%%%%%%%%%%%%%%%%%%%%%%%
We have identified the equations of motion for the open string tachyon
rolling in a linear dilaton background as being delay differential
equations. This class of equations has been widely studied in the
mathematics literature. Although no method to analytically solve
non-trivial DDEs is known, the method of steps is in general an
efficient numerical method (and essentially the only method) once the
initial data is provided. We have used this method in order to solve
the equations of motion at level zero, and found that it is both
efficient and accurate; in particular our solution agrees very well
with the results from an exponential series ansatz, and can be
continued much further in light-cone time (Fig.~\ref{fig:level0}).

At truncation level two, we found that the method of steps can be used
only when we further truncate the action by either removing all higher
derivatives, or by keeping only the scalar fields. But we saw that the
results from these approximations all agree reasonably well with the
more accurate solution from the exponential series ansatz, at least
until the first maximum has been reached
(Fig.~\ref{fig:CompTach}). The obtained numerical solutions can be
continued for much larger light-cone time, where it is seen that they
converge towards the non-perturbative vacuum. If we don't simplify the
equations of motion, our numerical solver gives us wrong solutions. We
have understood that this problem is due to the fact that our
equations of motion possess derivatives of order higher than one at
retarded times. In general, such equations are not even guaranteed to
possess continuous solutions. While we still expect the equations of
motion to have an analytic solution for the initial data corresponding
to a tachyon in the perturbative vacuum in the far light-cone past,
the method of steps is rendered unstable by the higher derivatives.

\begin{figure}
\begin{center}
\input{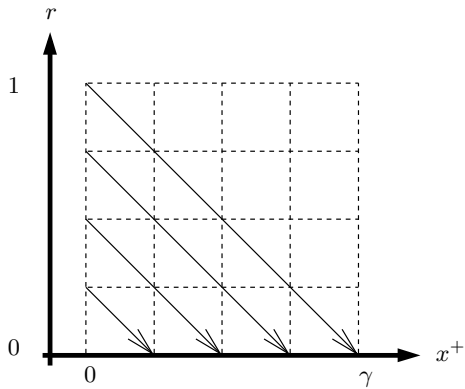}
\caption{\footnotesize{
Step  2 in the algorithm to solve a DDE in the diffusion equation approach. 
The initial data at $\xp = 0$ is copied ``diagonally'' to the $r=0$ boundary.}}
\label{fig_PDE_method}
\end{center}
\end{figure}

We have already mentioned that the diffusion equation method
\cite{Calcagni:2007wy,Calcagni:2009tx,Calcagni:2009jb} has been
successfully used to solve numerically the light-like rolling problem
at level zero \cite{Barnaby}. It is then natural to ask wether this
method can give more stable solutions at higher levels. The diffusion
method has the virtue of applying to a larger class of non-local
equations. But in the case of DDEs, we believe that it is completely
identical to the method of steps. It would then face exactly the same
problem as the method of steps at higher levels. To illustrate this,
we briefly review the diffusion method applied to the equation
\begin{equation}
\phi'(x^+) - \phi(x^+)=-\phi(x^+-\gamma)^2.
\label{exampleeq}
\end{equation} 
First we define a function of two variables $x^+$ and $r$ by
$
\Phi(x^+,r) \equiv e^{\gamma r \partial_+} \phi(x^+) =
\phi(x^+ +r\gamma).
$ This definition is implemented by the simple diffusion equation
\begin{equation}
\partial_+ \Phi(x^+,r) = \frac{1}{\gamma} \partial_r \Phi(x^+,r).
\label{diffusion}
\end{equation}
The idea is then to solve the diffusion equation on the region $\xp \ge 0$
 and $0\leq r\leq1$. The subtle part is to determine the boundary
conditions. It turns out that the boundary conditions on the $r=1$
axis are determined by the equation (which is now local in $x^+$):
\begin{equation}
\Phi(x^+,1)-\frac{1}{\gamma} \left[ \partial_r
   \Phi(x^+,r)\right]_{r=1} = \Phi(x^+,0)^2.
\label{boundary}
\end{equation}
A simple algorithmic resolution of the diffusion equation is the following:
\begin{enumerate}
\item Specify boundary conditions on the segment $x^+ = 0$. This is the {\em initial data}.
\item Solve the diffusion equation (\ref{diffusion}) with the given piece of
 boundary conditions. One can easily see that this amounts to copying ``diagonally''
 the values on the segment $x^+=0$ to the horizontal segment defined by $r=0$ and
 $0 \leq x^+ \leq \gamma$, see Fig.~\ref{fig_PDE_method}. \label{diagonally}
\item To solve the diffusion equation further, one needs first to specify the 
initial conditions on the segment defined by $r=1$ and $0 \leq x^+ \leq \gamma$ 
using Eq.~(\ref{boundary}). Since the simple diffusion equation has been solved 
exactly in the previous step,  replace $\frac{\partial_r}{\gamma}$ by
 $\partial_+$, and the right-hand side is simply the initial data. One 
obtains precisely the original equation (\ref{exampleeq}) (with $\phi(x^+)$ 
replaced by $\Phi(x^+,1)$).
\item Solve again the diffusion equation by copying the obtained boundary
 values on the segment $r=1$ and $0 \leq x^+ \leq \gamma$ to the segment $r=0$
 and $\gamma \leq x^+ \leq 2 \gamma$.
\item Solve again Eq.~(\ref{boundary}) on the segment $r=1$ and
 $\gamma \leq x^+ \leq 2 \gamma$. From the same argument as above, plus the 
fact that one can replace the right-hand side of (\ref{boundary}) by 
$\Phi(x^+ -\gamma, 1)^2$, this is solving the DDE equation (\ref{exampleeq})
 on the interval $\gamma \leq x^+ \leq 2 \gamma$ given the values of $\phi$ 
on the interval $0 \leq x^+ \leq \gamma$. This is precisely the method of steps
 as described in Appendix \ref{s_dde}!
\end{enumerate}
The algorithm then continues by repeating points 4 and 5, which is
equivalent to solving the original DDE \eqref{exampleeq} by the method
of steps.

To summarize, the diffusion method applied to a DDE ``mimics'' the method of steps. We cautiously note, however, that a different numerical method to solve the diffusion equation may change the above claim if Eq.~(\ref{diffusion}) is {\em not} solved exactly as in our algorithm. 

\paragraph{}
In order to find accurate numerical solutions taking all derivatives
into account, we had to resort to the exponential series ansatz. This
method generalizes obviously from level zero; in particular one still
can calculate recursively the coefficients of the series. The problem
with this method is that the number of coefficients needed grows
exponentially with the light-cone time $x_\text{max}^+$ until which we
want our solution to be accurate. Moreover, we must calculate the
coefficients with many digits precision in order to account for the
numerical cancellation errors. At levels (2,4) (Figs.~\ref{fig:ExpAll}
and \ref{fig:ExpTachyon}) and (2,6) (Fig.~\ref{fig:ExpAll26}) we were
nevertheless able to find solutions that are accurate up to a point
where the string field is convincingly seen to converge towards the
non-perturbative vacuum.  Moreover, we have found that the convergence
(as measured by the overshooting of the first maximum) is
substantially faster at levels two and four than at level zero. At
level four, it is a little bit slower that at level two
(Fig.~\ref{fig:lev2vs4}). In order to compare this to the analytic
solution of Hellerman and Schnabl \cite{Hel-Sch}, it is instructive to
look at the tachyon component of their solution (in Schnabl
gauge). Normalizing the tachyon vev in the non-perturbative vacuum to
one, it is
\begin{equation}
f_1^{(0)}(x^+) = \frac{e^{x^+}}{1+e^{x^+}},
\label{HStach}
\end{equation}
a monotonic solution. So the fact that our solution at level two has
become more monotonic is a good indication that level truncation
yields qualitatively the same solution (in Siegel gauge). We note,
however, that if we expand (\ref{HStach}) as a power series in
$e^{x^+}$ (what we have called an exponential series), it has a finite
radius of convergence; the series diverges for $x^+ \geq 0$. On the
other hand, the exponential series that we obtained have a much larger
radius of convergence, perhaps infinite as can be shown at level zero
from the asymptotic behavior of the coefficients of the series
\cite{Hel-Sch}. It is worth mentioning a curious fact. Had we taken
the level-zero equation of motion (with vev normalized to unity)
$\phi'(\xp) - \phi(\xp) = - \phi(\xp-\gamma)^2$, and {\em dropped the
  delay} $\gamma$, we would get
$$
\phi'(\xp) - \phi(\xp) = -\phi(\xp)^2,
$$
which has the solution
$$
\phi(x^+) = \frac{e^{x^+}}{1+e^{x^+}}.
$$ Precisely the same as (\ref{HStach})! This might just be a
coincidence; it could also point at something deeper which may be
worth examining in a further work.

\paragraph{}
To sum up, our numerical solutions at levels two and four strongly
indicate that the string field will eventually converge to the
non-perturbative vacuum, in agreement with the analytic solution of
\cite{Hel-Sch}. On the other hand, we have also shown that, at levels
higher than zero, the equations of motion, linearized around the
non-perturbative vacuum, admit infinitely many oscillating solutions
with exponentially growing amplitude. This suggests that the string
field should not converge to this vacuum.
These two apparently contradictory results can be reconciled in
different  ways. Below we describe two possibilities.

\paragraph{} 
The \textit{first} possibility is that the string field somehow avoids the
infinitely many exponentially growing modes. This is certainly
possible because the initial conditions (i.e. the string field sitting
in the perturbative vacuum in the infinite light-cone past) are very
special: 
With this point of view, we should conclude that the motion
is unstable: should an external perturbation, or a quantum effect,
move the string field the tiniest bit away from this particular
solution it would not converge. We should also stress that our
analysis of the linearized equations of motion doesn't tell us how
non-convergent the rolling would be. It might approach a limit cycle
(in phase space) as is sometimes the case when the linearized equations of
motion have a {\em finite} number of growing solutions. Our situation,
however, is different because we have infinitely many growing
solutions in the vacuum. This could render the rolling diverging. Or
it might not be that bad; after all, we could imagine expressing a
well-behaved converging function as a series in the infinitely many
diverging modes \cite{Bellman-Cooke}; this would not be possible in
the finite case. We emphasize here again the special character of the
initial condition of the tachyon sitting on the top of the hill in the
infinite light-cone past, which we have shown uniquely fixes the
initial data up to a shift in light-cone time. The method of
exponential series, which gives a convergent solution, can work only
with this particular initial condition, which is also the initial
condition of the analytic solution of \cite{Hel-Sch}. For a generic
initial condition the coefficients cannot be determined recursively,
because ``negative'' modes have to be taken into account. In the level
zero case, with $\phi(x^+) = \sum_{n=-\infty}^{\infty}a_n e^{n x^+}$,
one obtains
\begin{equation}
 a_n = -\frac{K^{3-2n} }{n-1} \sum_{m=-\infty}^{\infty}{a_m a_{n-m}}.
\label{recursion relation}
\end{equation}
In other words: in order to calculate \textit{one} coefficient, we need 
to know \textit{all} coefficients, including the one we are interested in!  
The same problem is found in the DDE approach: to find initial data that
are consistent with the equation of motion, we need to solve the very
same equation of motion. This is a simple problem only when we require
that the tachyon be in the non-perturbative vacuum in the infinite
past. 

We leave it for further work to study different initial conditions. At
level zero it was done in \cite{Barnaby}; but at higher levels we
expect different results due to the higher derivatives and divergent
modes.

\paragraph{} 
A \textit{second} possibility to reconcile a convergent rolling
solution with infinitely many growing modes is that these modes do not
satisfy the ``out-of-gauge'' equations of motion\footnote{We thank
  T.~Erler and M.~Schnabl for emphasizing this point to us.}. Indeed
we have derived the equations of motion from the gauge-fixed action.
This means that for a solution $\Psi_0$, we are guaranteed that 
$\langle \Phi, Q \Psi_0 + \Psi_0 \star \Psi_0 \rangle = 0$ for all $\Phi$ in 
Siegel gauge. But it is possible that $Q \Psi_0 + \Psi_0 \star \Psi_0$ itself doesn't 
vanish. That such solutions $\Psi_0$ can exist was shown in \cite{Hata:2000bj}.

\paragraph{} 
The two different explanations above differ in particular by the
interpretation of the growing modes. In the first possibility, they do
have some physical meaning and make the rolling unstable. While we do
not expect to find open string degrees of freedom in the
non-perturbative vacuum, it is not unreasonable to think that the
unstable modes could correspond to the {\em closed} tachyon
instability. After all, we have coupled the open string field theory
to the closed sector, although only via the dilaton. In the second
possibility mentioned above, however, the growing modes are simply
unphysical because they do not obey the out-of-gauge equations of
motion. In order to find out if this is the case, we would have to
derive the equations of motion from the gauge-unfixed action and
verify whether the growing modes are still solutions of these full
equations of motion. We hope to be able to report on this in a future
publication.  The present paper gives only a partial answer to the
question of the physical meaning of the growing modes. At the end of
Section \ref{s_oscillations}, we wrote an effective equation of motion
for one scalar field, whose solutions are precisely the exponential
modes solutions of the whole system of equations of motion linearized
around the non-perturbative vacuum. We have then found that the
decaying modes correspond to {\em positive} delays in the effective
equation of motion, thus non-locality in the {\em past}, and that
growing modes correspond to {\em negative} delays, i.e. 
non-locality extending into the {\em future}.

\acknowledgments
We extend our sincere gratitude to T.~Erler for useful discussions, and to 
N.~Barnaby and M.~Schnabl for insightful 
comments on the manuscript. We are indebted to A.~Golovnev for providing 
us with a copy of Ref.~\cite{Ostrogradski}. N.~M. is supported in parts
by the Transregio TRR 33 'The Dark Universe' and the Excellence
Cluster 'Origin and Structure of the Universe' of the DFG.

%%%%%%%%%%%%%%%%%%%%%%%%%%%%%%%%%%%%%%%%%%%%%%%%%%%%%%%%%%%%%%%%
\appendix
%%%%%%%%%%%%%%%%%%%%%%%%%%%%%%%%%%%%%%%%%%%%%%%%%%%%%%%%%%%%%%%%

%%%%%%%%%%%%%%%%%%%%%%%%%%%%%%%%%%%%%%%%%%%%%%%%%%%%%%%%%%%%%%%%
\section{Delay differential equations}
\label{s_dde}
Differential equations describing real life systems like chemical
reactions, predator-prey relations, etc... often need to incorporate
the state of the system not only now, but also at times in the past. A
simple example is the following {\em delay differential equation} (DDE)
\begin{equation}
y'\left(t\right)=-y\left(t-1\right) \label{eq: retarded DDE}
\end{equation}
with the \emph{constant delay} term $y\left(t-1\right)$. Notice the
non-locality: the derivative is needed at $t$, but the function itself
is evaluated a finite distance away at $t-1$. This equation is of the
\textit{retarded} type. In order to make this a well posed problem it
is \emph{not} sufficient to look for a solution starting at a point
$t_{0}$ with the value of the function $y\left(t_{0}\right)=y_{0}$
fixed as in the Cauchy initial value problem for an ordinary
differential equation (ODE). Instead one has to supply the\emph{
  initial data} $\phi\left(t\right)$ over a continuous range of
points. Choosing for example $\phi\left(t\right)=1$ one has as a well
posed problem
$$
\begin{cases}
y'\left(t\right)=-y\left(t-1\right) & t>0\\
y\left(t\right)=1 & t\le0.
\end{cases}
$$
In contrast to an ODE problem the solution to the above problem has
\emph{discontinuity points} of order \emph{n }where the $n$-th derivative
is discontinuous. One can quickly see how this comes about. Approaching
0 from the left, $\lim_{t\to0^{-}}y'\left(t\right)=0$, however approaching
from the right one finds
$$
\lim_{t\to0^{+}}y'\left(0\right)=-y\left(-1\right)=-1.
$$
This discontinuity propagates further along the solution. The second
derivative at $t=1$ reads
$$
y''\left(1\right)=-y'\left(0\right),
$$
and is thus discontinuous. However, the first derivative at $t=1$ is
$y'\left(1\right)=-y\left(0\right)$, which is continuous. Hence the
solution is of class $C^{1}$ at $t=1.$ This process continues at every
positive integer $t=n$, where the solution is $C^{n}$; hence the
discontinuities are smoothed out as $t$ increases. 
 The propagation
and existence of discontinuity points is a generic feature of DDEs. In
general the solution only links continuously with the initial data at
the starting point $t_{0}$.

\begin{figure}
\begin{center}
\input{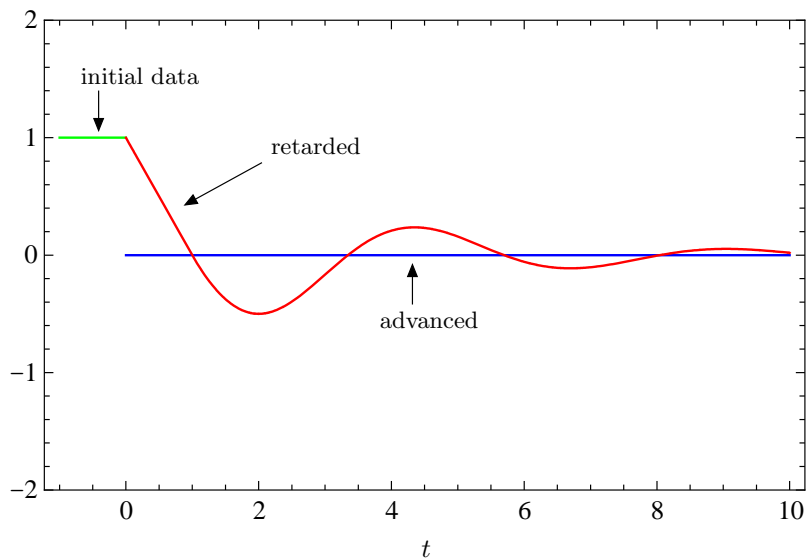}
\caption{\footnotesize{
Solutions to the retarded  and advanced examples with identical initial data.
 The retarded equation
possesses a unique continuous solution for any initial data, but the
advanced solution is in general discontinuous.}}
\label{fig: DDEavanced}
\end{center}
\end{figure}

Consider as a second example a DDE of the \textit{advanced} type
\begin{equation} 
\begin{cases}
y'\left(t\right)=-y\left(t+1\right) & t>0\\
y\left(t\right)=1 & t\le0 \quad. \label{eq: advanced DDE}
\end{cases}
\end{equation}
The only variation to \eqref{eq: retarded DDE} is that the derivative is
taken at the present instant, but the function is needed at a future point.
For $-1<t<0$ we have $y\left(t+1\right)=y'\left(t\right)=0$ and similarly
the solution vanishes for all positive $t$. The solutions to both examples,
\eqref{eq: retarded DDE} and \eqref{eq: advanced DDE}, are plotted in 
Figure \ref{fig: DDEavanced}. The discontinuity of the advanced solution at
$t=0$ is noteworthy: for this type of equation only special initial conditions
allow for a continuous solution. In our example that special choice among 
the class of constant initial data is  $y\left(t\right)=0,\quad t\le0$.

The above examples can be generalized further to include any number of
delays, possibly \emph{variable }or \emph{time dependent delays
}$\tau\left(t\right)$ or even \emph{state dependent delays}
$\tau\left(t,y\left(t\right)\right)$.  If there are derivative terms
at delayed points,
$y'\left(t-\sigma\left(t,y\left(t\right)\right)\right)$ the equation
is said to be of the \emph{neutral type (}NDDE). In general the delay
functions $\tau\left(t,y\left(t\right)\right)$ and
$\sigma\left(t,y\left(t\right)\right)$ can be completely independent.
If $f$ depends on $y$ over a continuous range, the problem is called a
\emph{Volterra functional differential equation}, but for the present
purpose, one can restrict to only a finite number of points in the
past required to compute the current rate of change in the system.

The general definition of a first-order delay differential equation
problem then reads
\begin{equation*}
\left\{ \begin{array}{lll}
y'(t) = & f\left(t,y(t), y(t-\tau_1(t,y(t))), \ldots 
y(t-\tau_m(t,y(t))) , \right. &  \\
& \quad \left.  y'(t-\sigma_1(t,y(t))), \ldots y'(t-\sigma_n(t,y(t))) \right)
& t_0 \le t \le t_f \\
y(t)= & \phi(t) & t \le t_0.
\end{array}
\right.
\end{equation*}
Extending to systems of DDEs or higher order derivative terms is then
straightforward. There are further major differences between ODEs
and DDEs in addition to the above discussed discontinuities.
\begin{itemize}
\item Different initial data $\phi\left(t\le t_{0}\right)$ can give rise
to the same solution $y\left(t\ge t_{0}\right)$
\item The solution may be non-unique in the state dependent case
\item The solution may terminate at a point
\item In the neutral case generally no smoothing of discontinuities occurs
\item Bounded solutions may oscillate or show chaotic behavior even in
  the one-dimensional case, while ODEs can oscillate only in at least
  two dimensions and behave chaotically in at least three dimensions
  (Poincar\'e-Bendixson theorem). In fact oscillations are typical of
  DDEs
\item infinitesimal delays introduced into ODEs can stabilize or destabilize
the solution
\item For scalar NDDEs no smoothing occurs, but for systems of NDDEs smoothing
may take place 
\end{itemize}

\paragraph*{Existence and Uniqueness}

Theorems on local existence and uniqueness have been developed for
very general assumptions on the delay functions $\tau\left(t,y\left(t\right)\right)$
and $\sigma\left(t,y\left(t\right)\right)$. But for the case at hand
in this paper one can use familiar results from ODE theory. Assume
there is only one delay, a positive number on the interval of interest
$\left[t_{0},t_{f}\right]$ \[
\tau=\sigma>0.\]
This case of constant delay then reduces to the following \emph{ordinary}
differential equation initial value problem on the interval $\left[t_{0},t_{0}+\tau\right]$
between the initial point and the corresponding point shifted by one
delay \[
\begin{cases}
y'\left(t\right)=f\left(t,y\left(t\right),\phi\left(t-\tau\right),\phi'\left(t-\tau\right)\right) & t_{0}\le t\le t_{0}+\tau\\
y\left(t_{0}\right)=\phi\left(t_{0}\right).\end{cases}\]
Standard results then guarantee local existence and uniqueness of
a continuous solution in $\left[t_{0},t_{0}+\delta\right]$ for some $\delta>0$
provided that $f\left(t,y\left(t\right),\phi\left(t-\tau\right),\phi'\left(t-\tau\right)\right)$
is continuous with respect to $t$, Lipschitz continuous with respect
to $y,\phi,\phi'$ and that $\phi\left(t\right),\phi'\left(t\right)$
are themselves Lipschitz continuous. Note however that there exists
a whole family of discontinuous solutions, linked to the unique
continuous one by a constant offset.

\paragraph*{Method of steps}

In order to obtain the solution on the full interval it is convenient
both theoretically and computationally to iterate along the intervals
$\left[t_{0}+j\cdot\tau,t_{0}+\left(j+1\right)\tau\right]$, $j=0,1,2,\dots$
using the solution calculated for the previous interval. This is known
as the \emph{method of steps}, the principal algorithm to solve DDEs
numerically. It is applicable in the case where all discontinuity
points are known in advance, e.g. for a constant delay. Solving the
DDE then reduces to solving ODEs on the intervals between two discontinuity
points using standard ODE methods such as \emph{linear multistep }solvers
and gluing the solution together at the interval boundaries. For continuing
over several such intervals the ODE method has to supply a continuous
(polynomial) interpolation of the solution which matches the accuracy
order of the solver. For concreteness, let $y_{0}$ be the solution
on the interval $\left[t_{0},t_{0}+\tau\right]$, then for the solution
$y_{1}$in the next step the ODE reads\[
\begin{cases}
y_{1}'\left(t\right)=f\left(t,y_{1}\left(t\right),y_{0}\left(t-\tau\right),y_{0}'\left(t-\tau\right)\right) & t_{0}+\tau\le t\le t_{0}+2\tau\\
y_{1}\left(t_{0}+\tau\right)=y_{0}\left(t_{0}+\tau\right).\end{cases}\]
Further references are, for example,
\cite{Bellman-Cooke}, \cite{Driver} and \cite{Bellen}.

%%%%%%%%%%%%%%%%%%%%%%%%%%%%%%%%%%%%%%%%%%%%%%%%%%%%%%%%%%%%%%%%

%%%%%%%%%%%%%%%%%%%%%%%%%%%%%%%%%%%%%%%%%%%%%%%%%%%%%%%%%%%%%%%%%%%%%%%%%%%%%%
\section{Lagrangian at level (2,4)}
\label{Lagrangian24}
%%%%%%%%%%%%%%%%%%%%%%%%%%%%%%%%%%%%%%%%%%%%%%%%%%%%%%%%%%%%%%%%%%%%%%%%%%%%%%

We present the full Lagrangian density with fields up to level two
and interaction terms up to level four in the linear dilaton background
with light-like dilaton gradient $V^{2}=0$.  The Lagrangian can
be split into two: one is explicitly $V$-dependent and vanishes for
$V=0$, the other survives for $V=0$. While $V$ could always be
eliminated using the modified momentum conservation $\sum k_{i}^{\mu}+iV^{\mu}=0$,
we keep it explicit for easier comparison with \cite{Kos-Sam}.
The interaction term can be split up according to the level $n$ and
the number of derivatives $k$, $\mathcal{L}_{k}^{\left(n\right)}$.
Overall we have \begin{eqnarray*}
S & = & -\frac{1}{g^{2}}\int\mbox{d}^{D}x\,\, e^{-V\cdot x}\left\{ \frac{1}{2}\left(\mathcal{L}_{Kin}+\mathcal{L}_{Kin}^{V}\right)+\frac{1}{3}K^{3}\left(\mathcal{L}_{Int}+\mathcal{L}_{Int}^{V}\right)\right\} \\
\mathcal{L}_{Int} & = & \mathcal{L}^{\left(0\right)}+\mathcal{L}^{\left(2\right)}+\mathcal{L}_{0}^{\left(4\right)}+\mathcal{L}_{1}^{\left(4\right)}+\mathcal{L}_{2}^{\left(4\right)}+\mathcal{L}_{3}^{\left(4\right)}+\mathcal{L}_{4}^{\left(4\right)}.\end{eqnarray*}
We use the following expansion of the string field into Fock space
states

\[
\state{\Psi}=\biggl\{\phi+\frac{i}{\sqrt{2}}B_{\mu}\alpha_{-2}^{\mu}+\frac{1}{\sqrt{2}}B_{\mu\nu}\alpha_{-1}^{\mu}\alpha_{-1}^{\nu}\,+\beta b_{-1}c_{-1}\biggr\} c_{1}\state 0.\]
Working in the light-cone, we 
have \[ e^{-V\cdot x}=e^{V^{+} x^{-}} \] and for example  \[V_{\mu} B^{\mu} = - V^{+}B^{-}. \]
We split the Lorentz indices into light-like and ordinary components
 \[ \mu=\left(0,1,2,\dots D-1\right)\to\left(+,-,2,3,\dots D-1\right)\equiv\left(+,-,i\right).\] 
In order to simplify the equations of motion, we need not consider all 379
independent components 
of  $B_{\mu \nu}, B_{\nu},\beta,\phi$ from the expansion (where $B_{\mu\nu}$ is symmetric).
 Several facts aid in reducing the number of fields:
\begin{enumerate} 
\item Any symmetric tensor can be reduced to a traceless and a pure trace part: 
$B_{ij}=S_{ij}+\frac{1}{24}F\eta_{ij}$ with $S_{ij}$ traceless.  
\item For any field 
$\Phi_{1}\in\Omega=\left\{ B^{i},\, B^{+i},\, B^{ij},\, B^{-i},\, S^{ij}\right\} $, 
it only appears in the action coupled to another field $\Phi_{2}\in\Omega$. 
Thus we can safely set these fields to zero and trivially satisfy their equations
 of motion, but still allow for a non-trivial solution for the remaining fields.  
\end{enumerate} We end up with the following eight fields to solve for: 
\[ \left\{ \phi,\, B^{+},\, B^{-},\, B^{++},\, B^{+-},\, B^{--},\, F,\,\beta\right\} .\]
Fields with a tilde are defined by \[
\tilde{\phi}=K^{\Box}\phi,\qquad K=\frac{3\sqrt{3}}{4}.\]
Explicitly the kinetic and interaction $\L$ without the extra dilaton
contribution contain the following 41 terms:
\footnote{For brevity we set $\alpha'=1$.
}

\newcommand{\ti}[1]{\tilde{#1}\,}

\begin{eqnarray*}
\mathcal{L}_{kin} & = & \partial_{\mu}\phi\partial^{\mu}\phi+\partial_{\mu}B_{\nu}\partial^{\mu}B^{\nu}+\partial_{\mu}B_{\nu\lambda}\partial^{\mu}B^{\nu\lambda}-\partial_{\mu}\beta\partial^{\mu}\beta\\
 &  & -\phi^{2}+B_{\mu}B^{\mu}+B_{\mu\nu}B^{\mu\nu}-\beta^{2}\\
 \mathcal{L}^{(2)} & = & \frac{-5}{9\sqrt{2}}\ti{B}_{\mu}\,^{\mu}\ti{\phi}^{2}-\frac{11}{\sqrt{3}}\ti{\beta}\ti{\phi}^{2}-\frac{2}{3}\ti{\phi}^{2}\partial_{\mu}\ti{B}^{\mu}-\frac{16}{9\sqrt{2}}\ti{\phi}\partial_{\mu}\partial_{\nu}\ti{\phi}\ti{B}^{\mu\nu}+\frac{16}{9\sqrt{2}}\partial_{\mu}\ti{\phi}\,\partial_{\nu}\ti{\phi}\,\ti{B}^{\mu\nu}\\
\mathcal{L}_{0}^{(4)} & = & \frac{128}{81}\ti{B}_{\mu}\ti{B}^{\mu}\ti{\phi}+\frac{25}{486}\ti{B}_{\mu}\,^{\mu}\ti{B}_{\nu}\,^{\nu}\ti{\phi}+\frac{512}{243}\ti{B}_{\mu\nu}\ti{B}^{\mu\nu}\ti{\phi}+\frac{\sqrt{2}\,\,55}{243}\ti{\beta}\ti{B}_{\mu}\,^{\mu}\ti{\phi}+\frac{19}{81}\ti{\beta}^{2}\ti{\phi}\\
\mathcal{L}_{1}^{(4)} & = & \frac{256\,\sqrt{2}}{243}\ti{\phi}\partial_{\mu}\ti{B}_{\nu}\ti{B}^{\mu\nu}-\frac{256\,\sqrt{2}}{243}\partial_{\mu}\ti{\phi}\ti{B}_{\nu}\ti{B}^{\mu\nu}+\frac{10\,\sqrt{2}}{81}\ti{\phi}\partial_{\mu}\ti{B}^{\mu}\ti{B}_{\nu}\,^{\nu}+\frac{44}{81}\ti{\phi}\ti{\beta}\partial_{\mu}\ti{B}^{\mu}\\
\mathcal{L}_{2}^{(4)} & = & \frac{40}{243}\ti{\phi}\ti{B}_{\mu\nu}\partial^{\mu}\partial^{\nu}\ti{B}_{\rho}\,^{\rho}+\frac{40}{243}\partial^{\mu}\partial^{\nu}\ti{\phi}\ti{B}_{\mu\nu}\ti{B}_{\rho}\,^{\rho}-\frac{80}{243}\partial^{\mu}\ti{\phi}\ti{B}_{\mu\nu}\partial^{\nu}\ti{B}_{\rho}\,^{\rho}\\
 &  & -\frac{512}{243}\partial_{\mu}\ti{\phi}\ti{B}_{\nu\lambda}\partial^{\nu}\ti{B}^{\mu\lambda}+\frac{512}{243}\partial_{\mu}\partial_{\nu}\ti{\phi}\ti{B}_{\lambda}\,^{\nu}\ti{B}^{\mu\lambda}+\frac{512}{243}\ti{\phi}\partial_{\mu}\ti{B}_{\nu\lambda}\partial^{\nu}\ti{B}^{\mu\lambda}\\
 &  & -\frac{176\sqrt{2}}{83}\partial_{\mu}\ti{\phi}\ti{B}^{\mu\nu}\partial_{\nu}\beta+\frac{88\sqrt{2}}{83}\ti{\phi}\ti{B}^{\mu\nu}\partial_{\mu}\partial_{\nu}\beta+\frac{88\sqrt{2}}{83}\partial_{mu}\partial_{\nu}\ti{\phi}\ti{B}^{\mu\nu}\beta\\
 &  & +\ti{\phi}\partial_{\mu}\ti{B}^{\mu}\partial_{\nu}\ti{B}^{\nu}\\
\mathcal{L}_{3}^{(4)} & = & -\frac{32\,\sqrt{2}}{81}\partial_{\mu}\ti{\phi}\partial_{\nu}\partial_{\lambda}\ti{B}^{\lambda}\ti{B}^{\mu\nu}+\frac{16\,\sqrt{2}}{81}\ti{\phi}\partial_{\mu}\partial_{\nu}\partial_{\lambda}\ti{B}^{\lambda}\ti{B}^{\mu\nu}+\frac{16\,\sqrt{2}}{81}\partial_{\mu}\partial_{\nu}\ti{\phi}\partial_{\lambda}\ti{B}^{\lambda}\ti{B}^{\mu\nu}\\
\mathcal{L}_{4}^{(4)} & = & \frac{32}{243}\ti{\phi}\partial_{\mu}\partial_{\nu}\ti{B}^{\rho\sigma}\partial_{\rho}\partial_{\sigma}\ti{B}^{\mu\nu}+\frac{64}{243}\partial_{\mu}\partial_{\nu}\ti{\phi}\ti{B}^{\rho\sigma}\partial_{\rho}\partial_{\sigma}\ti{B}^{\mu\nu}\\
 &  & \frac{32}{243}\partial_{\mu}\partial_{\nu}\partial_{\rho}\partial_{\sigma}\ti{\phi}\ti{B}^{\rho\sigma}\ti{B}^{\mu\nu}-\frac{128}{243}\ti{\phi}\partial_{\mu}\partial_{\nu}\partial_{\rho}\ti{B}^{\mu\sigma}\partial_{\sigma}\ti{B}^{\nu\rho}\\
 &  & -\frac{128}{243}\partial_{\mu}\ti{\phi}\partial_{\nu}\ti{B}^{\rho\sigma}\partial_{\rho}\partial_{\sigma}\ti{B}^{\mu\nu}+\frac{128}{243}\partial_{\mu}\partial_{\nu}\ti{\phi}\partial_{\rho}\ti{B}_{\sigma}\,^{\nu}\partial^{\sigma}\ti{B}^{\mu\rho}\end{eqnarray*}

In addition we obtain eight interaction terms and four kinetic terms from introducing the dilaton. The kinetic terms however do not contribute to the equations of motion, they are essentially total derivatives.

\begin{eqnarray*}
\mathcal{L}_{Kin}^{V} & = & V^{\nu} {B}^{\mu} \partial_{\nu}{B}_{\mu} + V^{\nu} {B}^{\mu\lambda} \partial_{\nu}{B}_{\mu\lambda} + V^{\nu}\phi \partial_{\nu} \phi -V^{\nu}\beta \partial_{\nu} \beta\\
\mathcal{L}_{Int}^{V} & = & \frac{11}{9} V_{\mu} \ti{B}^{\mu} \ti{\phi} \ti{\phi} + \frac{121}{243}  V_{\mu} \ti{B}^{\mu}  V_{\nu} \ti{B}^{\nu}  \ti{\phi} \\
 & & -\frac{44}{81}  V_{\mu} \ti{B}^{\mu}  \partial_{\nu} \ti{B}^{\nu}  \ti{\phi} - \frac{55 \sqrt{2}}{243} V_{\mu} \ti{B}^{\mu} \ti{B}_{\rho}\,^{\rho} \ti{\phi} - \frac{242}{243} V_{\mu} \ti{B}^{\mu} \ti{\beta} \ti{\phi} \\
 & & +\frac{88 \sqrt{2}}{243} V_{\mu} \left\{ 2  \partial_{\rho}  \ti{B}^{\mu} \ti{B}^{\rho\sigma} \partial_{\sigma} \ti{\phi} - \partial_{\rho}\partial_{\sigma} \ti{B}^{\mu} \ti{B}^{\rho\sigma}  \ti{\phi} - \ti{B}^{\mu} \ti{B}^{\rho\sigma}  \partial_{\rho}\partial_{\sigma}\ti{\phi}\right\} 
\end{eqnarray*}

\def\acomm#1#2{\left\{  #1,#2\right\}  }
\def\xm{X^{\mu}\left(z\right)}
\def\Vp{V^{+}}
\def\qvec{\overrightarrow{q}}
\def\del{\partial}
\def\corr#1{\left\langle #1\right\rangle }
\def\stateConj#1{\left\langle #1\right|}
\def\inpr#1#2{\left\langle #1\,\,\right|\left.#2\right\rangle }
\def\expectValue#1#2#3{\left\langle #1\right|#2\left|#3\right\rangle }
\def\normalOrdCreat{\genfrac{}{}{0pt}{2}{\circ}{\circ}}
\def\zb{\bar{z}}
\def\yp{y^{+}}
\def\pd{\left(2\pi\right)^{D}}
\def\brac#1{\left(#1\right)}
\def\xmu{X^{\mu}}
\def\delb{\bar{\partial}}
\def\comm#1#2{\left[#1,#2\right]}
\def\state#1{\left|#1\right\rangle }
\def\abl#1#2{\frac{\partial#1}{\partial#2}}
\def\L{\mathcal{L}}
\def\xnu{X^{\nu}}
\def\rt{\rightarrow}
\def\xp{x^{+}}
\def\dm{\partial_{-}}
\def\dpl{\partial_{+}}
\def\T{\tilde{\phi}}
 \def\B{\tilde{B}^{++}}
\def\F{\tilde{F}}
\def\G{\tilde{\beta}}
\def\P{\tilde{B}^{+-}}
\def\Q{\tilde{B}^{--}}
\def\R{\tilde{B}^{+}}
\def\S{\tilde{B}^{-}}
\def\TR{\left(-2\tilde{B}^{+-}+\F\right)}
\def\V{V^{+}}

\begin{align*}
 & \mbox{Writing the action explicitly in terms of these eight fields gives the Lagrangians:}\\
\mathcal{L}^{\left(2\right)}= & \frac{5}{9\sqrt{2}}\TR\T^{2}-\frac{11}{9}\G\T^{2}-\frac{2}{3}\sqrt{\alpha'}\left(\dpl\R+\dm\S\right)\T^{2}\\
 & -\frac{16}{9\sqrt{2}}\alpha'\T\left(\dpl^{2}\T\B+2\dpl\dm\T\P+\dm^{2}\T\Q\right)\\
 & +\frac{16}{9\sqrt{2}}\alpha'\left(\dpl\T\dpl\T\B+2\dpl\T\dm\T\P+\dm\T\dm\T\Q\right)\\
\mathcal{L}_{0}^{\left(4\right)}= & -2\cdot\frac{128}{81}\R\S\T+\frac{25}{486}\TR^{2}\T\\
&+\frac{256}{243}\T\left(2\B\Q+2\P\P+\frac{1}{24}\F^{2}\right) +\frac{\sqrt{2}\cdot55}{243}\G\T\TR+\frac{19}{81}\G^{2}\T\\
\mathcal{L}_{1}^{\left(4\right)}= & -\frac{256\cdot\sqrt{2}}{243}\sqrt{\alpha'}\T\left(\dpl\S\B+\left(\dpl\R+\dm\S\right)\P+\dm\R\Q\right)\\
 & +\frac{256\cdot\sqrt{2}}{243}\sqrt{\alpha'}\left(\dpl\T\S\B+\left(\dpl\T\R+\dm\T\S\right)\P+\dm\T\R\Q\right)\\
 & +\frac{10\sqrt{2}}{81}\sqrt{\alpha'}\T\left(\dpl\R+\dm\S\right)\TR+\frac{44}{81}\sqrt{\alpha'}\T\G\left(\dpl\R+\dm\S\right)\\
\mathcal{L}_{2}^{\left(4\right)}= & \frac{40}{243}\alpha'\T\left\{ \B\dpl^{2}\TR+2\P\dpl\dm\TR \right.\\
&\left.+\Q\dm^{2}\TR\right\} \\
 & +\frac{40}{243}\alpha'\TR\left(\dpl^{2}\T\B+2\dpl\dm\T\P+\dm^{2}\T\Q\right)\\
 & -\frac{40}{243}\alpha'\biggl\{\dpl\T\B\dpl\TR+\dm\T\P\dpl\TR\\
 & \,\,\,\,\,\,\,\,\,\,\,\,\,\,\,+\dpl\T\P\dm\TR+\dm\T\Q\dm\TR\biggr\}\\
 & +\frac{512}{243}\alpha'\dpl\T\left(\Q\dm\B+\P\dm\P+\P\dpl\B+\B\dpl\P\right)\\
 & +\frac{512}{243}\alpha'\dm\T\left(\Q\dm\P+\P\dm\Q+\P\dpl\P+\B\dpl\Q\right)\\
 & -\frac{512}{243}\alpha'\left(\dpl^{2}\T\B\P+\dpl\dm\T\B\Q+\dpl\dm\T\P\P+\dm^{2}\T\P\Q\right)\\
 & -\frac{512}{243}\alpha'\T\left(\dpl\Q\dm\B+\dpl\P\dm\P+\dpl\B\dpl\P+\dm\P\dm\Q\right)\\
 & -\frac{88\cdot\sqrt{2}}{243}\alpha'\left(\dpl\T\B\dpl\G+\dpl\T\P\dm\G+\dm\T\P\dpl\G+\dm\T\Q\dm\G\right)\\
 & +\frac{88\cdot\sqrt{2}}{243}\alpha'\T\left(\B\dpl^{2}\G+2\P\dpl\dm\G+\Q\dm^{2}\G\right)\\
 & +\frac{88\cdot\sqrt{2}}{243}\alpha'\G\left(\B\dpl^{2}\T+2\P\dpl\dm\T+\Q\dm^{2}\T\right)+\frac{4}{27}\alpha'\T\left(\dpl\R+\dm\S\right)^{2}\\
\mathcal{L}_{3}^{\left(4\right)}= & -\frac{32\sqrt{2}}{81}{\alpha'}^{3/2}\dpl\T\left(\dpl^{2}\R\B+\dpl\dm\S\B+\dpl\dm\R\P+\dm^{2}\S\P\right)\\
 & -\frac{32\sqrt{2}}{81}{\alpha'}^{3/2}\dm\T\left(\dpl^{2}\R\P+\dpl\dm\S\P+\dpl\dm\R\Q+\dm^{2}\S\Q\right)\\
 & +\frac{16\sqrt{2}}{81}{\alpha'}^{3/2}\T\biggl(\dpl^{3}\R\B+2\dpl^{2}\dm\R\P+2\dpl\dm^{2}\S\P\\
 & \,\,\,\,\,\,\,\,\,\,\,\,\,\,\,\,\,\,\,\,\,\,\,\,\,\,\,\,\,\,\,\,+\dpl^{2}\dm\S\B+\dpl\dm^{2}\R\Q+\dm^{3}\S\Q\\
 & +\frac{16\sqrt{2}}{81}{\alpha'}^{3/2}\left(\dpl\R+\dm\S\right)\left(\dpl^{2}\T\B+2\dpl\dm\T\P+\dm^{2}\T\Q\right)\end{align*}

\begin{align*}
\mathcal{L}_{4}^{\left(4\right)}= & \frac{32}{243}{\alpha'}^{2}\T\left\{ \left(\dpl^{2}\B\right)^{2}+\left(\dm^{2}\Q\right)^{2}+2\dpl^{2}\Q\dm^{2}\B\right\} \\
 & +\frac{128}{243}{\alpha'}^{2}\T\left\{ \dpl\dm\B\dpl^{2}\P+\dpl\dm\Q\dm^{2}\P+\left(\dpl\dm\P\right)^{2}\right\} \\
 & +\frac{64}{243}{\alpha'}^{2}\biggl\{\dpl^{2}\T\B\dpl^{2}\B+\dm^{2}\T\Q\dm^{2}\Q\\
 & \,\,\,\,\,\,\,\,\,\,\,\,\,\,\,\,\,\,\,\,\,\,\,+2\dpl\dm\T\left(\B\dpl^{2}\P+2\P\dpl\dm\P+\Q\dm^{2}\P\right)\biggr\}\\
 & +\frac{64}{243}{\alpha'}^{2}\left\{ 2\dpl^{2}\T\P\dpl\dm\B+2\dm^{2}\T\P\dpl\dm\Q+\dpl^{2}\T\Q\dm^{2}\B\right.\\
& \left. \qquad +\dm^{2}\T\B\dpl^{2}\Q\right\} \\
 & +\frac{64}{243}{\alpha'}^{2}\dpl^{2}\dm^{2}\T\B\Q+\frac{32}{243}{\alpha'}^{2}\biggl\{\dpl^{4}\T\left(\B\right)^{2}+\dm^{4}\T\left(\Q\right)^{2}\\
 & \,\,\,\,\,\,\,\,\,\,+4\left(\dpl^{3}\dm\T\B\P+\dpl^{2}\dm^{2}\T\left(\P\right)^{2}+\dpl\dm^{3}\T\P\Q\right)\biggr\}\\
 & -\frac{128}{243}{\alpha'}^{2}\left\{ \dpl^{3}\T\left(\B\dpl\B+\P\dm\B\right)\right.\\
&\left.\qquad+2\dpl^{2}\dm\T\left(\B\dpl\P+\P\dm\P\right)\right\} \\
 & -\frac{128}{243}{\alpha'}^{2}\left\{ +2\dpl\dm^{2}\T\left(\P\dpl\P+\Q\dm\P\right)\right.\\
&\left.\qquad+\dm^{3}\T\left(\P\dpl\Q+\Q\dm\Q\right)\right\} \\
 & -\frac{128}{243}{\alpha'}^{2}\left\{ +\dpl\dm^{2}\T\left(\B\dpl\Q+\P\dm\Q\right)\right.\\
&\left. \qquad+\dpl^{2}\dm\T\left(\P\dpl\B+\Q\dm\B\right)\right\} \\
 & -\frac{128}{243}{\alpha'}^{2}\left\{ \dpl\T\dpl\B\dpl^{2}\B+\dm\T\dm\Q\dm^{2}\Q+\dpl\T\dpl\Q\dpl^{2}\B\right\} \\
 & -\frac{128}{243}{\alpha'}^{2}\left\{ \dm\T\dm\B\dpl^{2}\Q+\dpl\T\left(\dm\B\dpl^{2}\P+\dm\Q\dm^{2}\P\right)\right\} \\
 & -\frac{128}{243}{\alpha'}^{2}\left\{ \dm\T\left(\dpl\B\dpl^{2}\P+\dpl\Q\dm^{2}\P\right)+2\dpl\T\dpl\P\dpl\dm\B\right\} \\
 & -\frac{128}{243}{\alpha'}^{2}\left\{ 2\dpl\T\dm\P\dpl\dm\P+2\dm\T\dpl\P\dpl\dm\P\right.\\
& \left. \qquad+2\dm\T\dm\P\dpl\dm\Q\right\} \\
 & +\frac{128}{243}{\alpha'}^{2}\left\{ \dpl^{2}\T\left(\dpl\B\right)^{2}+\dm^{2}\T\left(\dm\Q\right)^{2}+\dpl^{2}\T\left(\dm\P\right)^{2}
%\right. \left. \qquad
+\dm^{2}\T\left(\dpl\P\right)^{2}\right\} \\
 & +\frac{256}{243}{\alpha'}^{2}\dm^{2}\T\dpl\Q\dm\P+\frac{128}{243}{\alpha'}^{2}\biggl\{2\dpl^{2}\T\dpl\P\dm\B\\
 & \,\,\,\,\,\,\,\,\,\,\,\,\,+2\dpl\dm\T\left(\dpl\P\dpl\B+\dm\P\dm\Q+\dpl\Q\dm\B\right.\\
&\left.\qquad \qquad+\dpl\P\dm\P\right)\biggr\}\\
\\\mathcal{L}_{Int}^{V}= & -\frac{11}{9}\sqrt{\alpha'}\V\S\T^{2}-\frac{55\sqrt{2}}{243}\sqrt{\alpha'}\V\S\TR\T+\frac{242}{243}\sqrt{\alpha'}\V\S\G\T\\
 & +\frac{121}{243}\alpha'\left(\V\S\right)^{2}\T+\frac{44}{81}\alpha'\V\S\T\left(\dpl\R+\dm\S\right)\\
 & -\frac{176\sqrt{2}}{243}{\alpha'}^{3/2}\V\left(\dpl\S\B\dpl\T+\dpl\S\P\dm\T+\dm\S\P\dpl\T\right.\\
& \left. \qquad+\dm\S\Q\dm\T\right)\\
 & +\frac{88\sqrt{2}}{243}{\alpha'}^{3/2}\V\left(\dpl^{2}\S\B\T+2\dpl\dm\S\P\T+\dm^{2}\S\Q\T\right)\\
 & +\frac{88\sqrt{2}}{243}{\alpha'}^{3/2}\V\S\left(\B\dpl^{2}\T+2\P\dpl\dm\T+\Q\dm^{2}\T\right)\\
\\\mathcal{L}_{Kin}= & -\dpl\phi\dm\phi-\left(\dpl B^{++}\dm B^{--}+\dpl B^{--}\dm B^{++}+\frac{1}{24}\dpl F\right)\\
 & -\left(\dpl B^{+}\dm B^{-}+\dpl B^{-}\dm B^{+}\right)+\dpl\beta\dm\beta\\
 & +\frac{1}{2\,\alpha'}\left\{ -\phi^{2}+2B^{++}B^{--}+2\left(B^{+-}\right)^{2}+\frac{F^{2}}{24}-2B^{+}B^{-}-\beta^{2}\right\} \end{align*}

%%%%%%%%%%%%%%%%%%%%%%%%%%%%%%%%%%%%%%%%%%%%%%%%%%%%%%%%%%%%%%%%

%%%%%%%%%%%%%%%%%%%%%%%%%%%%%%%%%%%%%%%%%%%%%%%%%%%%%%%%%%%%%%%%%%%%%%%%%%%%%%
\section{Equations of motion at level (2,4)}
\label{EoMs24}
%%%%%%%%%%%%%%%%%%%%%%%%%%%%%%%%%%%%%%%%%%%%%%%%%%%%%%%%%%%%%%%%%%%%%%%%%%%%%%

With the solution $B^{++}\equiv 0$ and the retarded point $\yp=\xp-2\alpha'V^{+}\log\left(K\right)$, the seven remaining equations of motion become:

\paragraph{Equation of motion for the tachyon $\phi$ (48 terms):}
\begin{align*}
0=\ & {V^{+}} {\phi}'(x^{+})-\frac{{\phi}(x^{+})}{{{\alpha}'} }\\
\begin{split}
+K^3 \biggl\{&\frac{55}{729} \sqrt{2} F(y^{+}) {\beta}(y^{+})-\frac{50}{729} F(y^{+}) {B^{+-}}(y^{+})+\frac{10}{243} \sqrt{2} \sqrt{{{\alpha}'} } F(y^{+}) {B^{+}}'(y^{+})\\
    &-\frac{5}{27} \sqrt{2} F(y^{+}) {\phi}(y^{+})+\frac{139 F(y^{+})^2}{4374}-\frac{110}{729} \sqrt{2} {\beta}(y^{+}) {B^{+-}}(y^{+})\\
    &+\frac{44}{243} \sqrt{{{\alpha}'} } {\beta}(y^{+}) {B^{+}}'(y^{+})-\frac{22}{27} {\beta}(y^{+}) {\phi}(y^{+})+\frac{19}{243} {\beta}(y^{+})^2\\
    &-\frac{256}{729} \sqrt{2} \sqrt{{{\alpha}'} } {B^{+}}(y^{+}) {B^{+-}}'(y^{+})-\frac{572}{729} \sqrt{2} \sqrt{{{\alpha}'} } {B^{+-}}(y^{+}) {B^{+}}'(y^{+})\\
    &+\frac{10}{27} \sqrt{2} {B^{+-}}(y^{+}) {\phi}(y^{+})+\frac{562}{729} {B^{+-}}(y^{+})^2-\frac{4}{9} \sqrt{{{\alpha}'} } {\phi}(y^{+}) {B^{+}}'(y^{+})\\
    &+\frac{4}{81} {{\alpha}'} {B^{+}}'(y^{+})^2-\frac{256}{243} {B^{+}}(y^{+}) {B^{-}}(y^{+})+{\phi}(y^{+})^2\biggr\}\\ 
\end{split}\\
\begin{split}
+K^3 {V^{+}} \Bigl\{&\frac{160}{729} {{\alpha}'}  {B^{+-}}(y^{+}) F'(y^{+})+\frac{80}{729} {{\alpha}'}  F(y^{+}) {B^{+-}}'(y^{+})+\frac{55}{729} \sqrt{2} \sqrt{{{\alpha}'} } F(y^{+}) {B^{-}}(y^{+})\\
    &+\frac{352}{729} \sqrt{2} {{\alpha}'}  {B^{+-}}(y^{+}) {\beta}'(y^{+})+\frac{176}{729} \sqrt{2} {{\alpha}'}  {\beta}(y^{+}) {B^{+-}}'(y^{+})\\
&+\frac{242}{729} \sqrt{{{\alpha}'} } {\beta}(y^{+}) {B^{-}}(y^{+}) +\frac{32}{243} \sqrt{2} \sqrt{{{\alpha}'} }^3 {B^{+-}}'(y^{+}) {B^{+}}'(y^{+})\\
&-\frac{16}{27} \sqrt{2} {{\alpha}'}  {\phi}(y^{+}) {B^{+-}}'(y^{+}) -\frac{224}{81} {{\alpha}'}  {B^{+-}}(y^{+}) {B^{+-}}'(y^{+})\\ 
&+\frac{64}{243} \sqrt{2} \sqrt{{{\alpha}'} }^3 {B^{+-}}(y^{+}) {B^{+}}''(y^{+}) -\frac{122}{243} \sqrt{2} \sqrt{{{\alpha}'} } {B^{+-}}(y^{+}) {B^{-}}(y^{+})\\
&-\frac{32}{27} \sqrt{2} {{\alpha}'}  {B^{+-}}(y^{+}) {\phi}'(y^{+}) -\frac{256}{729} \sqrt{2} \sqrt{{{\alpha}'} } {B^{--}}(y^{+}) {B^{+}}(y^{+})\\
&+\frac{44}{243} {{\alpha}'}  {B^{-}}(y^{+}) {B^{+}}'(y^{+})-\frac{22}{27} \sqrt{{{\alpha}'} } {B^{-}}(y^{+}) {\phi}(y^{+})\Bigr\}\\
\end{split}\\
\begin{split}
+K^3 (V^{+})^2 \Bigl\{&\frac{40}{729} {{\alpha}'}  F(y^{+}) {B^{--}}(y^{+})+\frac{88}{729} \sqrt{2} {{\alpha}'}  {\beta}(y^{+}) {B^{--}}(y^{+})\\
        &+\frac{512}{729} {{\alpha}'} ^2 {B^{+-}}(y^{+}) {B^{+-}}''(y^{+})+\frac{176}{729} \sqrt{2} \sqrt{{{\alpha}'} }^3 {B^{-}}(y^{+}) {B^{+-}}'(y^{+})\\
        &+\frac{640}{729} {{\alpha}'} ^2 {B^{+-}}'(y^{+})^2-\frac{592}{729} {{\alpha}'}  {B^{+-}}(y^{+}) {B^{--}}(y^{+})\\
        &+\frac{352}{729} \sqrt{2} \sqrt{{{\alpha}'} }^3 {B^{+-}}(y^{+}) {B^{-}}'(y^{+})+\frac{16}{243} \sqrt{2} \sqrt{{{\alpha}'} }^3 {B^{--}}(y^{+}) {B^{+}}'(y^{+})\\
        &-\frac{8}{27} \sqrt{2} {{\alpha}'}  {B^{--}}(y^{+}) {\phi}(y^{+})+\frac{121}{729} {{\alpha}'}  {B^{-}}(y^{+})^2\Bigr\}\\ 
\end{split}\\
\begin{split}
+K^3 (V^{+})^3 \Bigl\{&\frac{128}{729} {{\alpha}'} ^2 {B^{--}}(y^{+}) {B^{+-}}'(y^{+})+\frac{256}{729} {{\alpha}'} ^2 {B^{+-}}(y^{+}) {B^{--}}'(y^{+})\\
            &+\frac{88}{729} \sqrt{2} \sqrt{{{\alpha}'} }^3 {B^{--}}(y^{+}) {B^{-}}(y^{+})\Bigr\}\\
\end{split}\\
\begin{split}
 +K^3 (V^{+})^4 \Bigl\{+\frac{32}{729}  {{{\alpha}'}} ^2 {B^{--}}(y^{+})^2\Bigr\}\\
\end{split}
\end{align*}

\paragraph{Equation of motion for $B^{++}$ (30 terms):}
\begin{align*}
0=\ &{V^{+}} {B^{--}}'(x^{+})+\frac{{B^{--}}(x^{+})}{{{\alpha}'} }\\
\begin{split}
+K^3 \Bigl\{&\frac{40}{729} {{\alpha}'}  {\phi}(y^{+}) F''(y^{+})-\frac{80}{729} {{\alpha}'}  F'(y^{+}) {\phi}'(y^{+})+\frac{40}{729} {{\alpha}'}  F(y^{+}) {\phi}''(y^{+})\\
    &+\frac{88}{729} \sqrt{2} {{\alpha}'}  {\phi}(y^{+}) {\beta}''(y^{+}) -\frac{176}{729} \sqrt{2} {{\alpha}'}  {\beta}'(y^{+}) {\phi}'(y^{+})+\frac{88}{729} \sqrt{2} {{\alpha}'}  {\beta}(y^{+}) {\phi}''(y^{+})\\
    &+\frac{16}{27} {{\alpha}'}  {\phi}(y^{+}) {B^{+-}}''(y^{+}) +\frac{224}{243} {{\alpha}'}  {B^{+-}}'(y^{+}) {\phi}'(y^{+})-\frac{368}{243} {{\alpha}'}  {B^{+-}}(y^{+}) {\phi}''(y^{+})\\
    &+\frac{512}{729} {B^{--}}(y^{+}) {\phi}(y^{+})+\frac{16}{243} \sqrt{2} \sqrt{{{\alpha}'} }^3 {B^{+}}^{(3)}(y^{+}) {\phi}(y^{+})\\
&-\frac{32}{243} \sqrt{2} \sqrt{{{\alpha}'} }^3 {B^{+}}''(y^{+}) {\phi}'(y^{+}) +\frac{16}{243} \sqrt{2} \sqrt{{{\alpha}'} }^3 {B^{+}}'(y^{+}) {\phi}''(y^{+})\\
&-\frac{256}{729} \sqrt{2} \sqrt{{{\alpha}'} } {\phi}(y^{+}) {B^{-}}'(y^{+}) +\frac{256}{729} \sqrt{2} \sqrt{{{\alpha}'} } {B^{-}}(y^{+}) {\phi}'(y^{+})\\
&-\frac{8}{27} \sqrt{2} {{\alpha}'}  {\phi}(y^{+}) {\phi}''(y^{+})+\frac{8}{27} \sqrt{2} {{\alpha}'}  {\phi}'(y^{+})^2\Bigr\}\\ 
\end{split}\\
\begin{split}
+K^3 {V^{+}} \Bigl\{&\frac{128}{729} {{\alpha}'} ^2 {B^{+-}}^{(3)}(y^{+}) {\phi}(y^{+})-\frac{128}{243} {{\alpha}'} ^2 {B^{+-}}'(y^{+}) {\phi}''(y^{+})\\
&+\frac{256}{729} {{\alpha}'} ^2 {B^{+-}}(y^{+}) {\phi}^{(3)}(y^{+}) +\frac{512}{729} {{\alpha}'}  {\phi}(y^{+}) {B^{--}}'(y^{+})\\
&-\frac{512}{729} {{\alpha}'}  {B^{--}}(y^{+}) {\phi}'(y^{+})+\frac{88}{729} \sqrt{2} \sqrt{{{\alpha}'} }^3 {\phi}(y^{+}) {B^{-}}''(y^{+})\\ 
&-\frac{176}{729} \sqrt{2} \sqrt{{{\alpha}'} }^3 {B^{-}}'(y^{+}) {\phi}'(y^{+}) +\frac{88}{729} \sqrt{2} \sqrt{{{\alpha}'} }^3 {B^{-}}(y^{+}) {\phi}''(y^{+})\Bigr\}\\ 
\end{split}\\
\begin{split}
+K^3 (V^{+})^2 \Bigl\{&\frac{64}{729} {{\alpha}'} ^2 {\phi}(y^{+}) {B^{--}}''(y^{+})-\frac{128}{729} {{\alpha}'} ^2 {B^{--}}'(y^{+}) {\phi}'(y^{+})\\
&+\frac{64}{729} {{\alpha}'} ^2 {B^{--}}(y^{+}) {\phi}''(y^{+})\Bigr\}\\
\end{split}
\end{align*}

\paragraph{Equation of motion for $B^{+-}$ (12 terms):}
\begin{align*}
0=\ &2 {V^{+}} {B^{+-}}'(x^{+})+\frac{2 {B^{+-}}(x^{+})}{{{\alpha}'} }\\
\begin{split}
  +K^3 \Bigl\{&-\frac{50}{729} F(y^{+}) {\phi}(y^{+})-\frac{110}{729} \sqrt{2} {\beta}(y^{+}) {\phi}(y^{+})+\frac{1124}{729} {B^{+-}}(y^{+}) {\phi}(y^{+})\\
        &-\frac{316}{729} \sqrt{2} \sqrt{{{\alpha}'} } {\phi}(y^{+}) {B^{+}}'(y^{+})+\frac{256}{729} \sqrt{2} \sqrt{{{\alpha}'} } {B^{+}}(y^{+}) {\phi}'(y^{+})+\frac{5}{27} \sqrt{2} {\phi}(y^{+})^2\Bigr\}\\
 \end{split}\\
\begin{split}
+K^3 {V^{+}} \Bigl\{&\frac{352}{729} {{\alpha}'}  {\phi}(y^{+}) {B^{+-}}'(y^{+})-\frac{832}{729} {{\alpha}'}  {B^{+-}}(y^{+}) {\phi}'(y^{+})\\
&-\frac{110}{729} \sqrt{2} \sqrt{{{\alpha}'} } {B^{-}}(y^{+}) {\phi}(y^{+})\Bigr\}\\
\end{split}\\
&+K^3 (V^{+})^2 \Bigl\{-\frac{80}{729}{{\alpha}'}  {B^{--}}(y^{+}) {\phi}(y^{+})\Bigr\}\\
\end{align*}

\paragraph{Equation of motion for $F$ (11 terms):}
\begin{align*}
0=\ &\frac{1}{24} {V^{+}} F'(x^{+})+\frac{F(x^{+})}{24 {{\alpha}'} }\\
\begin{split}
+K^3 \Bigl\{&\frac{139 F(y^{+}) {\phi}(y^{+})}{2187}+\frac{55}{729} \sqrt{2} {\beta}(y^{+}) {\phi}(y^{+})-\frac{50}{729} {B^{+-}}(y^{+}) {\phi}(y^{+})\\
    &+\frac{10}{243} \sqrt{2} \sqrt{{{\alpha}'} } {\phi}(y^{+})     {B^{+}}'(y^{+})-\frac{5 {\phi}(y^{+})^2}{27 \sqrt{2}}\Bigr\}\\ 
\end{split}\\
\begin{split}
+K^3 {V^{+}} \Bigl\{&\frac{80}{729} {{\alpha}'}  {\phi}(y^{+}) {B^{+-}}'(y^{+})+\frac{160}{729} {{\alpha}'}  {B^{+-}}(y^{+}) {\phi}'(y^{+})\\
&+\frac{55}{729} \sqrt{2} \sqrt{{{\alpha}'} } {B^{-}}(y^{+}) {\phi}(y^{+})\Bigr\}\\
\end{split}\\
&+K^3 (V^{+})^2\Bigl\{\frac{40}{729} {{\alpha}'}  {B^{--}}(y^{+}) {\phi}(y^{+}) \Bigr\}\\
\end{align*}

\paragraph{Equation of motion for $\beta$ (11 terms):}
\begin{align*}
0=\ &-{V^{+}} {\beta}'(x^{+})-\frac{{\beta}(x^{+})}{{{\alpha}'} }\\
\begin{split}
+K^3 \Bigl\{&\frac{55}{729} \sqrt{2} F(y^{+}) {\phi}(y^{+})+\frac{38}{243} {\beta}(y^{+}) {\phi}(y^{+})-\frac{110}{729} \sqrt{2} {B^{+-}}(y^{+}) {\phi}(y^{+})\\
        &+\frac{44}{243} \sqrt{{{\alpha}'} } {\phi}(y^{+}) {B^{+}}'(y^{+})-\frac{11}{27} {\phi}(y^{+})^2\Bigr\}\\ 
\end{split}\\
\begin{split}
+K^3 {V^{+}} \Bigl\{&\frac{176}{729} \sqrt{2} {{\alpha}'}  {\phi}(y^{+}) {B^{+-}}'(y^{+})+\frac{352}{729} \sqrt{2} {{\alpha}'}  {B^{+-}}(y^{+}) {\phi}'(y^{+})\\
&+\frac{242}{729} \sqrt{{{\alpha}'} } {B^{-}}(y^{+}) {\phi}(y^{+})\Bigr\}\\
\end{split}\\
&+K^3 (V^{+})^2 \Bigl\{\frac{88}{729} \sqrt{2} {{\alpha}'}  {B^{--}}(y^{+}) {\phi}(y^{+})\Bigr\}\\
\end{align*}

\paragraph{Equation of motion for $B^{+}$ (20 terms):}
\begin{align*}
0=\ &-{V^{+}} {B^{-}}'(x^{+})-\frac{{B^{-}}(x^{+})}{{{\alpha}'} }\\
\begin{split}
+K^3 \Bigl\{&-\frac{10}{243} \sqrt{2} \sqrt{{{\alpha}'} } {\phi}(y^{+}) F'(y^{+})-\frac{10}{243} \sqrt{2} \sqrt{{{\alpha}'} } F(y^{+}) {\phi}'(y^{+})-\frac{44}{243} \sqrt{{{\alpha}'} } {\phi}(y^{+}) {\beta}'(y^{+})\\
    &-\frac{44}{243} \sqrt{{{\alpha}'} } {\beta}(y^{+}) {\phi}'(y^{+})+\frac{316}{729} \sqrt{2} \sqrt{{{\alpha}'} } {\phi}(y^{+}) {B^{+-}}'(y^{+})\\
&+\frac{572}{729} \sqrt{2} \sqrt{{{\alpha}'} } {B^{+-}}(y^{+}) {\phi}'(y^{+})-\frac{8}{81} {{\alpha}'}  {\phi}(y^{+}) {B^{+}}''(y^{+})-\frac{8}{81} {{\alpha}'}  {B^{+}}'(y^{+}) {\phi}'(y^{+})\\
&-\frac{256}{243} {B^{-}}(y^{+}) {\phi}(y^{+})
    +\frac{4}{9} \sqrt{{{\alpha}'} } {\phi}(y^{+}) {\phi}'(y^{+})\Bigr\}\\ 
\end{split}\\
\begin{split}
+K^3 {V^{+}} \Bigl\{&-\frac{32}{243} \sqrt{2} \sqrt{{{\alpha}'} }^3 {\phi}(y^{+}) {B^{+-}}''(y^{+})-\frac{32}{81} \sqrt{2} \sqrt{{{\alpha}'} }^3 {B^{+-}}'(y^{+}) {\phi}'(y^{+})\\
        &-\frac{64}{243} \sqrt{2} \sqrt{{{\alpha}'} }^3 {B^{+-}}(y^{+}) {\phi}''(y^{+})+\frac{256}{729} \sqrt{2} \sqrt{{{\alpha}'} } {B^{--}}(y^{+}) {\phi}(y^{+})\\
        &-\frac{44}{243} {{\alpha}'}  {\phi}(y^{+}) {B^{-}}'(y^{+})-\frac{44}{243} {{\alpha}'}  {B^{-}}(y^{+}) {\phi}'(y^{+})\Bigr\}\\ 
\end{split}\\
&+K^3 (V^{+})^2 \Bigl\{-\frac{16}{243} \sqrt{2} \sqrt{{{\alpha}'} }^3 {\phi}(y^{+}) {B^{--}}'(y^{+})-\frac{16}{243} \sqrt{2} \sqrt{{{\alpha}'} }^3 {B^{--}}(y^{+}) {\phi}'(y^{+})\Bigr\}\\
\end{align*}

\paragraph{Equation of motion for $B^{-}$ (12 terms):}
\begin{align*}
0=\ &-{V^{+}} {B^{+}}'(x^{+})-\frac{{B^{+}}(x^{+})}{{{\alpha}'} }\\
& +K^3 \Bigl\{-\frac{256}{243} {B^{+}}(y^{+}) {\phi}(y^{+})\Bigr\}\\
\begin{split}
+K^3 {V^{+}} \Bigl\{&\frac{25}{729} \sqrt{2} \sqrt{{{\alpha}'} } F(y^{+}) {\phi}(y^{+})+\frac{110}{729} \sqrt{{{\alpha}'} } {\beta}(y^{+}) {\phi}(y^{+})+\frac{206}{729} \sqrt{2} \sqrt{{{\alpha}'} } {B^{+-}}(y^{+}) {\phi}(y^{+})\\
        &+\frac{20}{243} {{\alpha}'}  {\phi}(y^{+}) {B^{+}}'(y^{+})-\frac{5}{27} \sqrt{{{\alpha}'} } {\phi}(y^{+})^2\Bigr\}\\
\end{split}\\
\begin{split}
 +K^3 (V^{+})^2 \Bigl\{&\frac{80}{729} \sqrt{2} \sqrt{{{\alpha}'} }^3 {\phi}(y^{+}) {B^{+-}}'(y^{+})+\frac{160}{729} \sqrt{2} \sqrt{{{\alpha}'} }^3 {B^{+-}}(y^{+})  {\phi}'(y^{+})\\ &+\frac{110}{729} {{\alpha}'}  {B^{-}}(y^{+}) {\phi}(y^{+})\Bigr\}\\
\end{split}\\
&+K^3 (V^{+})^3 \Bigl\{\frac{40}{729}  \sqrt{2}\sqrt{{{\alpha}'} }^3 {B^{--}}(y^{+}) {\phi}(y^{+})\Bigr\}
\end{align*}

%%%%%%%%%%%%%%%%%%%%%%%%%%%%%%%%%%%%%%%%%%%%%%%%%%%%%%%%%%%%%%%%%%%%%%%%%%%%%%
\section{Determinant of polynomial matrices}
\label{polydet_s}
%%%%%%%%%%%%%%%%%%%%%%%%%%%%%%%%%%%%%%%%%%%%%%%%%%%%%%%%%%%%%%%%%%%%%%%%%%%%%%

In this paper, we have to evaluate the determinant of a matrix $M$
whose entries are polynomials in two variables $\omega$ and $y$. The
result is obviously a polynomial in $\omega$ and $y$ as well. One
could of course use the traditional methods, for example LU
decomposition, but one then notices that the complexity of the
calculation grows very fast with the matrix size. This is easy to
understand when one notes that such methods require the matrix to be
over a field. One will then have to think of elements of $M$ as
rational functions of $\omega$ and $y$. The calculation of the
determinant will then involve potentially heavy computations on large
fractions. Moreover, since the determinant is a polynomial, the end
result will involve a large simplification between numerator and
denominator, meaning that we might be doing many more calculations
than necessary. In fact we observed that, while \texttt{mathematica}
can calculate such a determinant for a seven by seven matrix, it fails
to do so for a fifty by fifty matrix whose entries are polynomials of
degrees at most eight in $\omega$ and at most one in $y$. This
prompted us to look for alternative methods.

An elegant method \cite{Sebek1} is based on the discrete Fourier
transform. We mention also that there is another method \cite{Sebek2},
that can efficiently calculate the degree of the determinant. This is
in fact what we need, but it is unclear to us how to generalize this
method to polynomial matrices in several variables, so we will use the
first method based on the Fourier transform.

To explain this method, it will be enough to concentrate on the case
of polynomials in one variable $x$; the generalization to several
variables will then be obvious.  The crucial fact to realize is that
polynomial multiplication is, in some sense, a convolution. More
explicitly, let's consider two polynomials $p(x)$ and $q(x)$ of
degrees $d_1$ and $d_2$
\begin{equation}
p(x) = \sum_{n=0}^{d_1} p_n \, x^n \quad \text{and} \quad 
q(x) = \sum_{m=0}^{d_2} q_m \, x^m.
\end{equation}
We define $N = d_1 + d_2 + 1$ and we write the coefficients $p_n$ and
$q_n$ into vectors ${\mathbf p}$ and ${\mathbf q}$, both of length
$N$.
\begin{equation}
{\mathbf p} = (p_0, p_1, \ldots, p_{N-1}) \quad , \quad 
{\mathbf q} = (q_0, q_1, \ldots, q_{N-1}),
\end{equation}
where $p_n = 0$ for $n>d_1$ and $q_m = 0$ for $m>d_2$. We can then
write the product $t(x) = p(x) q(x)$ also in a vector of the same
length because it has degree $N-1$. The components of ${\mathbf t}$
are then given by $t_n = \sum_{m=0}^{n} p_m q_{n-m}$. If we define the
negative components of ${\mathbf p}$ and ${\mathbf q}$ by imposing the
periodicity $p_n = p_{N+n}$ (and similarly for $q_n$ and $t_n$), we
can extend the summation to the whole range
\begin{equation}
t_n = \sum_{m=0}^{N-1} p_m q_{n-m}.
\label{cyclic_convolution}
\end{equation}
This in fact doesn't change the sum because all additional terms are
zero; but in the form (\ref{cyclic_convolution}), one recognizes that
${\mathbf t}$ is the {\em cyclic convolution} of ${\mathbf p}$ and
${\mathbf q}$
\begin{equation}
{\mathbf t} = {\mathbf p} * {\mathbf q}.
\end{equation}
We can now use the fact that the discrete Fourier transform defined by
\begin{equation}
{\cal F}({\mathbf a}) = (A_0, A_1, \ldots, A_{N-1}), \quad \text{with} \quad
A_k = \sum_{n=0}^{N-1} e^{2 \pi i k n / N} a_n.
\end{equation}
changes cyclic convolution into multiplication. Namely
\begin{equation}
{\cal F}({\mathbf t}) = {\cal F}({\mathbf p}) \, {\cal F}({\mathbf q}),
\end{equation}
where the multiplication of two vectors is done component-wise. One
can then perform an inverse discrete Fourier transform to obtain
$t(x)$.

Let us now consider, for a little while, the Leibniz formula for the
determinant $d(x)$ of the square matrix $M(x)$ of size $n$
\begin{equation}
d(x) = \sum_{\sigma \in S_n} \sgn(\sigma) \prod_{i=1}^n M_{i,\sigma(i)}(x).
\label{Leibniz}
\end{equation}
We estimate an upper bound $N-1$ on the degree of $d(x)$ and write
its coefficients in the vector ${\mathbf d}$ of length $N$; and we
also write the coefficients of $M_{i,j}(x)$ in vectors ${\mathbf
  M}_{i,j}$, each of length $N$. We can now write the polynomial
product in (\ref{Leibniz}) as a usual product of the components of the
discrete Fourier transforms of ${\mathbf M}_{i,j}$
\begin{equation}
{\cal F}({\mathbf d}) = \sum_{\sigma \in S_n} \sgn(\sigma) 
\prod_{i=1}^n {\cal F}({\mathbf M}_{i,\sigma(i)}).
\end{equation}
Now if we define the matrices ${\cal M}_n$ by
\begin{equation}
\left({\cal M}_n \right)_{i,j} \equiv ({\cal F}({\mathbf M}_{i,j}))_n,
\end{equation}
the Leibniz formula becomes
\begin{equation}
\left({\cal F}({\mathbf d})\right)_n = \sum_{\sigma \in S_n} \sgn(\sigma) \prod_{i=1}^n
({\cal M}_n)_{i,\sigma(i)}.
\end{equation}
In other words, the components of ${\cal F}({\mathbf d})$ are the
determinants of the matrices ${\cal M}_n$ whose entries are
numbers; these determinants can therefore be calculated with a
standard method. To summarize, the algorithm to calculate the
determinant $d(x)$ of $M(x)$ is:
\begin{enumerate}
\item Find an upper bound $N-1$ on the degree of $d(x)$.
\item Calculate the matrices ${\cal M}_n$ for $n=0,\ldots,N-1$, whose
  entries are the components $n$ of the discrete Fourier transform of
  the entries of ${\mathbf M}$.
\item Calculate the determinants of the matrices ${\cal M}_n$. These
  are the components of ${\cal F}({\mathbf d})$.
\item Inverse discrete Fourier transform ${\cal F}({\mathbf d})$ in order to
  find the coefficients of the polynomial $d(x)$.
\end{enumerate}

We make a few comments. First, it is well known that the discrete
Fourier transform can be made extremely fast with the Fast Fourier
Transform algorithm, so this method is relatively efficient. On the
other hand, if one is interested in determining the degree of $d(x)$,
numerical errors can make it hard to decide whether a term $\epsilon
\, x^n$ appearing in the answer for the determinant, with $\epsilon$ a
very small number in absolute value, should be kept or whether it
comes from numerical imprecision. For this reason, it might be
necessary to run the algorithm with a precision of many digits. At
last, this algorithm generalizes almost trivially to polynomials in
several variables; the discrete Fourier transforms have then to be
replaced by multidimensional discrete Fourier transforms.

%%%%%%%%%%%%%%%%%%%%%%%%%%%%%%%%%%%%%%%%%%%%%%%%%%%%%%%%%%%%%%%%%%%%%%%%%%%%%%
\section{Review of linear dilaton CFT}
\label{rev lin Dil}

%%%%%%%%%%%%%%%%%%%%%%%%%%%%%%%%%%%%%%%%%%%%%%%%%%%%%%%%%%%%%%%%%%%%%%%%%%%%%%

In the following we want to review briefly how
the linear dilaton CFT arises and quote the most useful results that
are needed to compute correlators and the equations of motion in a linear dilaton
background. A similar, even shorter overview can be found in \cite{Hel-Sch}, 
Section 2. Much more is covered in \cite{Polchinski:1998rq},
Sections 2.5, 2.7 and 3.7, where additionally the role 
of the dilaton field in string interactions is illuminated.

Let's start by considering the description of string theory in curved
spacetime. After all string theory is supposed to give a framework
for quantum gravity. As a reminder the Polyakov action
that serves as the starting point for string theory is 
\[
S_{P}=-\frac{1}{4\pi\alpha'}\int_{M}\mbox{d}^2\sigma\sqrt{-\gamma}\,\,\gamma^{\,\, ab}\partial_{a}X^{\mu}\partial_{b}X^{\nu}\eta_{\mu\nu}.\]
As a natural extension one is tempted to replace the flat Minkowski
metric $\eta_{\mu\nu}$ with a general metric $G_{\mu\nu}\left(X\right)$.
Considering spacetime as a coherent background of gravitons is slightly
dubious, as it means string theory is formulated in a background of
strings. But let's try anyway. If we add the graviton background we
might as well add the background from the two other closed string
massless states, the antisymmetric $B_{\mu\nu}$ tensor and the dilaton
$\Phi$, resulting in a so called nonlinear sigma model\index{nonlinear sigma model}
action\[
S_{\sigma}=-\frac{1}{4\pi\alpha'}\int_{M}\mbox{d}^2\sigma\sqrt{-\gamma}\,\,\left\{ \left(\gamma^{\,\, ab}G_{\mu\nu}\left(X\right)+i\epsilon^{ab}B_{\mu\nu}\left(X\right)\right)\partial_{a}X^{\mu}\partial_{b}X^{\nu}+\alpha'R\Phi\left(X\right)\right\}, \]
where $\gamma^{\,\, ab},R$ are the worldsheet metric and curvature. 

If we now demand this action to be Weyl invariant also at the quantum
level the corresponding $\beta$-functions for $G,B,\Phi$ of the
renormalization group flow have to vanish ($\Leftrightarrow$scale
invariance). To first order in $\alpha'$, only derivatives of the
dilaton $\Phi$ appear in the $\beta$-functions. This means changing
the dilaton by a constant is allowed.
 A constant
shift in $\Phi$ yields a contribution proportional to
 the  worldsheet Euler number $\chi$, defined as
\[ \chi = -\frac{1}{4\pi\alpha'}\int_{M}\mbox{d}^2\sigma\sqrt{-\gamma}R.  \]
In fact the Euler number determines the interaction strength, such that 
the open string coupling constant $g$ is related to the expectation 
value of the dilaton by
\[g^{2}=e^{\corr{\Phi}}.\]
Typically in quantum field theory the coupling is a free parameter, so different
values for the coupling represent different theories in sharp contrast
to string theory: different coupling is just a different background
of the same theory. However since we cannot determine these background
values from the dynamics, this distinction moves the difficulty elsewhere,
but in principle, there is no free parameter as in QFT. 

If the dilaton is not constant, then in order to preserve Weyl invariance,
the critical dimension of the string theory may be altered. One choice
of fields that preserves Weyl invariance is \[
G_{\mu\nu}\left(X\right)=\eta_{\mu\nu},\,\,\,\, B_{\mu\nu}=0,\,\,\,\,\Phi\left(X\right)=V_{\mu}X^{\mu},\]
where $\Phi$ is linear in $X$ (linear dilaton), and the constant
vector $V_{\mu}$ is the \emph{dilaton gradient. }It singles out a
particular direction in spacetime, breaking spacetime translation
symmetry. Momentum conservation is then modified as \begin{equation}
\sum_{j}k_{j}^{\mu}=0\to\sum_{j}k_{j}^{\mu}+iV^{\mu}=0\ .\label{eq:lin. dil. momentum conservation}\end{equation}
 The $\beta$ functions vanish if \[
V_{\mu}V^{\mu}=\frac{26-D}{6\alpha'}\ .\]
Conversely the central charge of this theory is \[
c=D+6\alpha'V_{\mu}V^{\mu}.\]
Hence different values of the dilaton gradient $V$ give different
CFTs. We will choose a light-like dilaton gradient, $V_{\mu}V^{\mu}=0$,
then the central charge and critical dimension are unaffected. 

The complete (Euclidean) action for the linear dilaton on worldsheet
$\Sigma$ with boundary $\del\Sigma$ then becomes\begin{equation}
S_{WS}=\frac{1}{4\pi\alpha'}\int_{\Sigma}\mbox{d}^{2}\sigma\sqrt{\gamma}\,\,\left(\gamma^{\,\, ab}\eta_{\mu\nu}\partial_{a}X^{\mu}\partial_{b}X^{\nu}+\alpha'RV_{\mu}X^{\mu}\right)+\frac{1}{2\pi}\int_{\del\Sigma}\mbox{d}s kV_{\mu}X^{\mu},\label{eq:lin. dilaton action}\end{equation}
where $k$ denotes the worldsheet boundary curvature. Varying
the action with respect to the metric gives the energy-momentum tensor
\begin{equation}
T\left(z\right)=-\frac{1}{\alpha'}:\del X_{\mu}\del X^{\mu}:+V_{\mu}\del^{2}X^{\mu},\label{eq:linear dilaton energy-momentum tensor}\end{equation}
the usual expression with an additional $V^{\mu}$ dependent contribution.

Gauge fixing the metric permits obtaining the same metric as before
on the complex plane, $\mbox{d}s^{2}=\mbox{d}z\mbox{d}\zb$ such that
the gauge fixed action has the same form as in the free theory with
$V_{\mu}=0$:\begin{equation}
S_{WS}=\frac{1}{2\pi\alpha'}\int\mbox{d}^{2}z\,\del X^{\mu}\delb X_{\mu}. \label{eq:gauge fixed}\end{equation}
This theory deserves the name CFT, for the action \eqref{eq:gauge fixed}
is invariant under the  conformal change of coordinate $z \to f(z)$,
where the field transforms as 
\begin{equation}
X^{\mu}\left(z,\zb\right)\to f\circ X^{\mu}\left(z,\zb\right)=X^{\mu}\left(f\left(z\right),\overline{f\left(z\right)}\right)+\frac{\alpha'}{2}V^{\mu}\log\left|f'\left(z\right)\right|^{2}.\label{eq:lin dil conformal trafo}\end{equation}
Note that this transformation has again an extra piece containing
the dilaton gradient. We want the (real)
boundary of the upper half plane to be mapped into itself so we can
use the doubling trick and $X^{\mu}\left(z,\zb\right)\to X^{\mu}\left(z\right)$,  
thus we require $\overline{f\left(z\right)}=f\left(\zb\right)$.
Denoting the real coordinate with $y\in\mathbb{R}$ in contrast to
complex $z$, a boundary field transforms as\[
X^{\mu}\left(y\right)\to X^{\mu}\left(f\left(y\right)\right)+\alpha'V^{\mu}\log\left|f'\left(y\right)\right|,\]
with the visible difference that the $2$s have disappeared. Expanding
the energy-momentum tensor into Laurent modes, we find the new (matter)
Virasoros\begin{equation}
L_{n}^{m}=\frac{1}{2}\sum_{k=-\infty}^{\infty}:\alpha_{n-k}^{\mu}\alpha_{\mu,k}:+i\sqrt{\frac{\alpha'}{2}}\left(n+1\right)V^{\mu}\alpha_{\mu,n}\ .
\label{eq:lin dil Virasoros}\end{equation}
The conformal weights of any operator can be determined from the OPE with
the energy-momentum tensor.
 For example, the weight of
$\left. :e^{ik\cdot X\left(z,\zb\right)}:\right.$ is \[
\alpha'\left(\frac{k^{2}}{4}+i\frac{V_{\mu}k^{\mu}}{2}\right).\]
On the boundary, things are a little more complicated. The normal ordering
\[:X^{\mu}\left(z_{1},\zb_{1}\right)X^{\nu}\left(z_{2},\zb_{2}\right):\ \equiv X^{\mu}\left(z_{1},\zb_{1}\right)X^{\nu}\left(z_{2},\zb_{2}\right)+\frac{\alpha'}{2}\eta^{\mu\nu}\log\left|z_{1}-z_{2}\right|^{2}\]
 is just valid in the bulk, not on the boundary. On the real line
the Green's function $G_{12}=-\frac{\alpha'}{2}\log\left|z_{1}-z_{2}\right|^{2}$
 needs to be modified (call it $G'_{12}$) to satisfy the Neumann
boundary conditions $\left(\del-\delb\right)G'_{12}=0$. This is accomplished
by an additional image charge term\begin{equation}
G'_{12}=-\frac{\alpha'}{2}\log\left|z_{1}-z_{2}\right|^{2}-\frac{\alpha'}{2}\log\left|z_{1}-\zb_{2}\right|^{2}.\label{eq:disk Green's function}\end{equation}
Normal ordering should remove the divergence of this Green's function
at zero separation, but the usual subtraction $\frac{\alpha'}{2}\eta^{\mu\nu}\log\left|z_{1}-z_{2}\right|^{2}$
is not enough, it only cancels the first divergent term, thus it has
to be doubled. The new prescription goes by the name of \emph{boundary
normal ordering,} distinguished from the bulk conformal normal ordering
by new symbols $\ordB\dots\ordB$,\[
\ordB X^{\mu}\left(y_{1}\right)X^{\nu}\left(y_{2}\right)\ordB\equiv X^{\mu}\left(y_{1}\right)X^{\nu}\left(y_{2}\right)+2\alpha'\log\left|y_{1}-y_{2}\right|.\]
We stress again that the reason for introducing another normal ordering
is just to have well-behaved (finite) expectation values for products
of boundary normal ordered operators. 

When taking real derivatives of boundary operators, factors of 2 need
to be taken into account because $\del_{z}=\frac{1}{2}\del_{y}$. This is a source of great confusion.
Furthermore operators on the boundary may
also have different weights from their {}``relatives'' defined on
the interior/bulk because of that, e.g. $\ordB e^{ik\cdot X\left(z\right)}\ordB$
has weight \begin{equation}
\alpha'\left(k^{2}+ik\cdot V\right).\label{eq:weight. of exp. on boundary}\end{equation}
It is instructive to check the consistency: the linear dilaton is
a background field, thus non-dynamical. If we set $V_{\mu}=0$ in
the above formulae the free string theory is recovered, as it should. 

%%%%%%%%%%%%%%%%%%%%%%%%%%%%%%%%%%%%%%%%%%%%%%%%%%%%%%%%%%%%%%%%%%%%%%%%%%%%%%

%%%%%%%%%%%%%%%%%%%%%%%%%%%%%%%%%%%%%%%%%%%%%%%%%%%%%%%%%%%%%%%%%%%%%%%%%%%%%%

\end{document}